\documentclass[nofootinbib,superscriptaddress,a4paper, twocolumn, floatfix]{revtex4-2}
\usepackage[english]{babel}
\usepackage[utf8]{inputenc}
\usepackage[colorinlistoftodos, color=green!40, prependcaption]{todonotes}
\usepackage{amsthm}
\usepackage{mathtools}
\usepackage{physics}
\usepackage{braket}
\usepackage{listings}
\usepackage{xcolor}
\usepackage{graphicx}
\usepackage{amssymb}
\usepackage[left=23mm,right=13mm,top=35mm,columnsep=15pt]{geometry} 
\usepackage{adjustbox}
\usepackage{placeins}
\usepackage[T1]{fontenc}
\usepackage{lipsum}
\usepackage{csquotes}
\usepackage[pdftex, pdftitle={Article}, pdfauthor={Author}]{hyperref} 
\usepackage{booktabs}
\usepackage{amssymb}
\usepackage{bm}
\usepackage{booktabs}
\usepackage[ruled,vlined]{algorithm2e}
\usepackage[caption=false]{subfig}

\newtheorem{statement}{Statement}
\begin{document}

\title{Unbiased observable estimation with approximate channels in fault-tolerant quantum computation}

\author{Dmitrii Khitrin}
\email[E-mail:]{dmitrii.khitrin@duke.edu}
\affiliation{Department of Physics, Duke University, Durham, NC 27708, USA.}
\affiliation{Duke Quantum Center, Duke University, Durham, NC 27701, USA.}

\author{Kenneth R. Brown}
\email[E-mail:]{kenneth.r.brown@duke.edu}
\affiliation{Department of Physics, Duke University, Durham, NC 27708, USA.}
\affiliation{Duke Quantum Center, Duke University, Durham, NC 27701, USA.}
\affiliation{Department of Electrical and Computer Engineering, Duke University, Durham, NC 27708, USA.}
\affiliation{Department of Chemistry, Duke University, Durham, NC 27708, USA.}

\author{Abhinav Anand}
\email[E-mail:]{abhinav.anand@duke.edu}
\affiliation{Duke Quantum Center, Duke University, Durham, NC 27701, USA.}
\affiliation{Department of Electrical and Computer Engineering, Duke University, Durham, NC 27708, USA.}

\date{\today}

\begin{abstract}
Unitary errors, such as those arising from fault-tolerant compilation of quantum algorithms, systematically bias observable estimates. 
Correcting this bias typically requires additional resources, such as an increased number of non-Clifford gates.
In this work, we present an alternative method for correcting bias in the expectation values of observables.
The method leverages a decomposition of the ideal quantum channel into a probabilistic mixture of noisy quantum channels. 
Using this decomposition, we construct unbiased estimators as weighted sums of expectation values obtained from the noisy channels.
We provide a detailed analysis of the method, identify the conditions under which it is effective, and validate its performance through numerical simulations. 
In particular, we demonstrate unbiased observable estimation in the presence of unitary errors by simulating the time dynamics of the Ising Hamiltonian.
Our strategy offers a resource-efficient way to reduce the impact of unitary errors, improving methods for estimating observables in noisy near-term quantum devices and fault-tolerant implementation of quantum algorithms.
\end{abstract}

\maketitle

\section{Introduction}
The past decade has seen remarkable advancements in the experimental realization of quantum computers~\cite{acharya2024quantum, bluvstein2024logical, kim2023evidence, moses2023race}. Significant progress has been made in building quantum devices with an increasing number of qubits~\cite{manetsch2024tweezer}, improved coherence times~\cite{wang2021single}, and higher gate fidelities~\cite{da2024demonstration}. Despite these achievements, current quantum devices remain susceptible to various noise sources, such as coherent and incoherent errors~\cite{unruh1995maintaining, landauer1995quantum, landauer1996physical, knill1998resilient, greenbaum2017modeling, gottesman1997stabilizer}, which limit their reliability and scalability. Addressing these challenges is essential for unlocking the full potential of quantum computing~\cite{548464, preskill1998reliable}.

In parallel, there has been considerable progress in developing algorithms tailored for quantum devices of the near-term~\cite{bharti2022noisy, cerezo2021variational}, early~\cite{katabarwa2024early, nelson2024assessment, anand2024stabilizer} and future~\cite{dalzell2023quantum, gottesman2010introduction, steane1999efficient, preskill1998fault} fault-tolerant era.
Near-term quantum algorithms~\cite{bharti2022noisy, cerezo2021variational}, such as those based on the variational quantum eigensolver (VQE)~\cite{peruzzo2014variational} and quantum approximate optimization algorithm (QAOA)~\cite{farhi2014quantum}, have been proposed for solving many problems under hardware constraints.
At the same time, advances in fault-tolerant protocols~\cite{leverrier2015quantum, panteleev2022asymptotically, breuckmann2021quantum} aim to enable error-free computation in the long term.
A central task in both these contexts is the accurate estimation of expectation values of observables.
Significant progress has been made to develop efficient subroutines for this task, including designing shorter circuits~\cite{anand2025hamiltonian, anand2023leveraging, tang2021qubit, grimsley2019adaptive, ryabinkin2018qubit, ryabinkin2020iterative, kandala2017hardware, anand2022quantum}, improving optimization procedures~\cite{schuld2019evaluating, kottmann2021feasible, izmaylov2021analytic, wierichs2022general, stokes2020quantum, sweke2020stochastic, anand2021natural, anand2024information}, reducing measurement complexity~\cite{huggins2021efficient, zhao2020measurement, gokhale2020n, izmaylov2019unitary, huang2020predicting}, optimizing  $T$-gate counts~\cite{ross2014optimal, heyfron2018efficient, kliuchnikov2023shorter, kissinger2019reducing, vandaele2024lower}, and developing new algorithms~\cite{bharti2022noisy, dalzell2023quantum}. 
However, noise in quantum devices restricts the effectiveness of these methods.


Several strategies have been proposed to address coherent and incoherent errors in quantum devices. These approaches span both quantum error correction \cite{548464, gottesman1997stabilizer, steane1999efficient, calderbank1996good, bravyi2018correcting, pato2024logical, beale2018quantum, huang2019performance}, and error mitigation \cite{BravyiGambetta, zhang2022hidden, cai2020mitigating, cai2023quantum, brown2004arbitrarily}. However, coherent errors, which take the form of unintended unitary operations due to systematic gate imperfections, remain particularly challenging to address due to their structured and persistent nature~\cite{sanders2015bounding, gutierrez2015comparison, kueng2016comparing, greenbaum2017modeling, iyer2018small}.
One common strategy for addressing these unitary errors involves composite pulses~\cite{brown2004arbitrarily, levitt1986composite, merrill2014progress, low2014optimal}, which are sequences of control operations specifically designed to counteract systematic imperfections and enhance the robustness of quantum gates.

In this work, we introduce an alternative method within the circuit model to suppress the effects of unitary errors on the measurement of an observable.
Consider a single-qubit Pauli rotation $R(\theta) = e^{-i\frac{\theta}{2}\mathbb{P}}$, where $\mathbb{P} \in \{\mathbb{X}, \mathbb{Y}, \mathbb{Z}\}$, and let $U'R(\theta)$ be the gate with a unitary error, $U'$.
Our aim is to mitigate the resulting bias in observable estimates due to $U'$ while keeping the fault-tolerant overhead minimal, meaning we do not decrease $\zeta$ and we introduce only a small number of additional $T$ gates.

If $U'$ is generated by the same Pauli operator $\mathbb{P}$, recent approaches~\cite{Koczor2022, Koczor2023, kiumi2024te} allow one to decompose the channel associated with $R(\theta)$ into a mixture of channels corresponding to $U'R(\theta)$, $U'R(\theta+\pi)$, and $U'R(\theta+\frac{\pi}{4})$, enabling the implementation of $R(\theta)$ using only noisy gates.
The idea of expressing an exact quantum channel as a quasi-probabilistic mixture of noisy ones was pioneered in Ref.~\cite{BravyiGambetta} and recently applied in Refs.~\cite{Koczor2023, kiumi2024te} to simulate quantum circuits using restricted gate sets.
In contrast, if $U'$ represents a rotation in an orthogonal plane, Pauli twirling techniques \cite{bennett1996mixed, dankert2009exact, emerson2007symmetrized, twirling_ex} can be employed to suppress its effect.
In this work, we combine these strategies, relaxing previous constraints on the structure of $U'$ allowing us to work with general unitary errors.

 Our  approach assumes prior knowledge of the unitary error channel,
 thus limiting its applicability to suppressing the effects of coherent errors in the fault-tolerant (FT) regime due to approximate gate synthesis.
In the fault-tolerant (FT) regime, unitary errors originating from approximate gate synthesis are known exactly. 
This makes it possible to decompose the channel for the exact circuit into a linear combination of noisy ones with appropriate coefficients and then construct an unbiased estimator for the expectation values of observables.
If the error arises from imperfections in physical gate implementations and given some information about the channel, our method can also be used to suppress the effect of such errors.
However, in principle, this  can  be used to recalibrate the control parameters and recover the ideal gate.

We demonstrate the utility of our method in both near-term and FT regime by performing various numerical simulation.
These results confirm that our method effectively mitigates the effects of unitary errors, including approximation-induced unitary errors in FT implementations and constant over-rotations in near-term devices.

The rest of the paper is organized as follows:
We begin with a review of  main results and preliminary theory in Sec.~\ref{sec:summary} and Sec.~\ref{sec:prelims}, respectively. The method and results from various numerical experiments are discussed in Sec.~\ref{sec:method} and Sec.~\ref{sec:results}, respectively.
Lastly, we provide concluding remarks in Sec.~\ref{sec:conclusion}.

\section{Summary and Main Results}\label{sec:summary}

Here, we state the problem we aim to solve, summarize the method we developed, and outline its key limitation.

\noindent\textbf{Problem statement.} Consider a quantum circuit $U$ composed of $\nu$ Pauli rotations:
\begin{equation}
    U = \prod_{i=1}^\nu R_i(\theta_i),
\end{equation}
where $R_i(\theta_i) \equiv e^{-i\frac{\theta_i}{2}\mathbb{P}_i}$ with $\mathbb{P}_i$ being a Pauli-string.
Given a finite-gate-set (e.g., Clifford+$T$) decomposition of each gate, the corresponding fault-tolerant circuit is
\begin{equation}
    \widetilde{U} = \prod_{i=1}^\nu U'_i R_i(\theta_i).
\end{equation}
Our goal is to estimate the expectation value $\Tr[\mathcal{O}\mathcal{U}\ketbra{0}{0}]$
of a Hermitian operator $\mathcal{O}$, while only executing $\widetilde{U}$ and its modified versions obtained by inserting a few additional Clifford or $T$ gates. Here, $\mathcal{U}(.) \equiv U(.)U^\dagger$ denotes the quantum channel corresponding to $U$ and the errors $\{U'_i\}$ are known in advance due to decomposition accuracy.

\noindent\textbf{Method:} Our strategy works by expressing $\mathcal{U}$ as a linear combination of noisy channels, namely $\widetilde{\mathcal{U}}$ and its slightly modified versions. 
This allows us to construct an unbiased estimator for $\Tr[\mathcal{O}\mathcal{U}\ketbra{0}{0}]$, mitigating the bias in the noisy estimate $\Tr[\mathcal{O}\widetilde{\mathcal{U}}\ketbra{0}{0}]$. 
The mixture of noisy channels we construct accounts for both the components of $U'_i$ along the $\mathbb{P}_i$ direction and those in the orthogonal direction. 
The former is achieved through strategies similar to Refs.~\cite{Koczor2022, Koczor2023, kiumi2024te}, while the latter is handled using Pauli twirling~\cite{bennett1996mixed, dankert2009exact, emerson2007symmetrized, twirling_ex}. 
The derivations of the channel decompositions and unbiased estimators are detailed in Sec.~\ref{sec:method}.

\noindent\textbf{Main Results:} Our method allows us to get accurate expectation values in the case when the accuracy of gate decomposition is relatively low ($\mathcal{O}(10^{-2})$) with circuits of moderate sizes ($\mathcal{O}(10^{3})$ gates) with very low T-gate count.
The reason behind this is the exponential sampling overhead in estimating the expectation value accurately. 
This overhead grows rapidly with both the number of gates $\nu$ and the error strength at each gate, as summarized in Statement~\ref{statement}. Such scaling limits the method's applicability to large circuits. Nevertheless, it remains valuable for reducing $T$-count in early fault-tolerant regimes.

\section{Preliminaries}\label{sec:prelims}
In this section, we review some of the essential theory required for the rest of the paper.

\subsection{Approximate unitary synthesis}\label{subsec:aus}
A single-qubit unitary can be expressed as:
\begin{equation}
    U = e^{-i\frac{\theta}{2}R},
\end{equation}
where $R= \eta_x \mathbb{X} + \eta_y \mathbb{Y} + \eta_z \mathbb{Z}$, $\{\eta_x, \eta_y, \eta_z\}$ are real numbers, $\{\mathbb{X}, \mathbb{Y}, \mathbb{Z}\}$ are the Pauli operators, and $\theta \in [0, 2\pi)$.
The channel induced by the unitary $U$ is given by:
\begin{equation}
    \mathcal{U} (.) = U (.) U^{\dagger}.
\end{equation}

The problem of approximating a single-qubit unitary~\cite{kitaev1997quantum, dawson2005solovay, ross2014optimal, kliuchnikov2023shorter}, $U$, within accuracy, $\zeta$, can be formulated as finding a unitary $V$, such that:
\begin{equation}
    \| U - V \| < \zeta.
\end{equation} 
The approximate unitary, $V$, can be written as $V = U'U$, such that
\begin{equation}
    U' = e^{-i\epsilon R'}
\label{eq:approx_unit_err}
\end{equation}
with $R'= \eta'_x \mathbb{X} + \eta'_y \mathbb{Y} + \eta'_z \mathbb{Z}$, $\epsilon \in [ 0, 2\pi)$ and $\| \mathbb{I} - U'\| < \zeta$.

\subsection{Coherent error}
Coherent errors are systematic deviations that arise due to imperfections in the implementation of quantum gates and classical control. 
These imperfections are unitary, and so the noisy gates that are implemented on the device can be written as the product of the ideal gate and a unitary error as follows:
\begin{equation}
    \widetilde{U} = U' U,
\end{equation}
where $\widetilde{U}$ is the noisy implemented gate, $U$ is the ideal gate and $U'$ is the coherent error.
The corresponding noisy channel is $\widetilde{\mathcal{U}} (.) = \widetilde{U} (.) \widetilde{U}^{\dagger}$.
One can also write this as the product of the ideal channel and the coherent error channel as:
\begin{equation}
    \widetilde{\mathcal{U}} (.) = \Lambda \mathcal{U}(.) = U' U (.) U^\dagger U'^{\dagger},
\end{equation}
where $\mathcal{U}$ is the ideal channel and $\Lambda$ is the coherent error channel.

A common form of coherent error is over(under)-rotation, i.e., given an ideal unitary $U = R(\theta)=e^{-i\frac{\theta}{2}\mathbb{P}}$, the noisy gate is of the form:
\begin{equation}
    \widetilde{U} = {R}(\theta+\epsilon)=e^{-i(\frac{\theta+\epsilon}{2})\mathbb{P}},
\end{equation}
where $\epsilon$ is some real number and $\mathbb{P} \in \left\{\mathbb{I}, \mathbb{X}, \mathbb{Y}, \mathbb{Z}\right\}^{\otimes n}$ is a Pauli-string.
The corresponding coherent error channel can be written as:
\begin{equation}\label{eq:cohe_err}
    \Lambda (.) = e^{-i\frac{\epsilon}{2}\mathbb{P}} (.) e^{i\frac{\epsilon}{2}\mathbb{P}}.
\end{equation}

\subsection{Pauli twirling}
Given a unitary $U$, the Pauli twirling operation over a subgroup, $\mathcal{G}$, of the Pauli group, is defined as:
\begin{equation}
    \mathcal{T}_{{U}} (.) = \frac{1}{|\mathcal{G}|} \sum_{\sigma\in\mathcal{G}} \sigma U \sigma^{\dagger} (.) \sigma^{\dagger} U^{\dagger} \sigma,
\end{equation}
where $|\mathcal{G}|$ is the cardinality of $\mathcal{G}$. Pauli twirling transforms coherent noise into a Pauli channel~\cite{bennett1996mixed, dankert2009exact, emerson2007symmetrized}, which is generally less detrimental to quantum computations. An application of this technique for accurately estimating observables is demonstrated in Ref.~\cite{twirling_ex}.

\subsection{Decomposition of gates and channels}
A unitary gate, $U = R(\theta)=e^{-i\frac{\theta}{2}\mathbb{P}}$, can be decomposed into a linear combination of two unitary gates (see Appendix\ref{ap:un_decomp} for details) as follows:
\begin{equation}
    R(\theta) = \cos(\frac{\epsilon}{2}) R(\theta+\epsilon) - \sin(\frac{\epsilon}{2})R(\theta+\epsilon+\pi),
\end{equation}
where $\epsilon \in [ 0, 2\pi)$. 

The corresponding channel, $\mathcal{U}=\mathcal{R}(\theta)$, can then be written (see Appendix\ref{ap:chan_decomp} for details) as follows:
\begin{align}\label{eq:full_decom}
    &\mathcal{R}(\theta) = \cos^{2}(\frac{\epsilon}{2}) \mathcal{R}(\theta+\epsilon) + \sin^{2}(\frac{\epsilon}{2}) \mathcal{R}(\theta+\epsilon+\pi) \nonumber\\
    &- \cos(\frac{\epsilon}{2}) \sin(\frac{\epsilon}{2})[\mathcal{R}\left(\theta+\epsilon+\frac{\pi}{2}\right) - \mathcal{R}\left(\theta+\epsilon-\frac{\pi}{2}\right)].
\end{align}

\section{Method}\label{sec:method}
In this section, we present the details of the proposed method and analyze the associated cost to estimate different properties of interest.

\subsection{Ideal channel as a mixture of noisy channels}
Given an ideal unitary gate $U = R(\theta) = e^{-i\frac{\theta}{2}\mathbb{P}}$, where $\mathbb{P} \in \left\{\mathbb{I}, \mathbb{X}, \mathbb{Y}, \mathbb{Z}\right\}^{\otimes N}$ is a Pauli-string and $\theta \in [0,2\pi)$.
We want to find a decomposition of this ideal gate in term of noisy gates, $\{\widetilde{U}_i = \widetilde{R}(\theta_i) = e^{-i(\frac{\theta_i+\epsilon}{2})\mathbb{P}}\}$, with $\epsilon$ being an over-rotation angle.

This problem can be formalized as finding the following channel decomposition:
\begin{equation}
\mathcal{R}(\theta) = \gamma_1\widetilde{\mathcal{R}}(\theta_1) + \gamma_2\widetilde{\mathcal{R}}(\theta_2) + \gamma_3\widetilde{\mathcal{R}}(\theta_3),
\label{eq:mixture}
\end{equation}
where $\mathcal{R}$ is the ideal channel, $\widetilde{\mathcal{R}}$ is the noisy channel and $\gamma_1, \gamma_2, \gamma_3$ are some functions of $\epsilon$.
This decomposition can allow us to perform the ideal unitary operation $\mathcal{R}$ while only having access to the noisy operations $\{\widetilde{\mathcal{R}}_i\}$.

By using Eq.~\ref{eq:full_decom} with Eq.~\ref{eq:mixture} and solving a system of equations (see Appendix\ref{ap:mix_chan} and~\ref{ap:opt_sol} for details), we find a decomposition:
\begin{align}
&\mathcal{R}(\theta) = \gamma_1\widetilde{\mathcal{R}}(\theta) + \gamma_2\widetilde{\mathcal{R}}(\theta-\frac{\pi}{4}) + \gamma_3\widetilde{\mathcal{R}}(\theta+\pi) \nonumber \\
&= \gamma_1\mathcal{R}(\theta+\epsilon) + \gamma_2\mathcal{R}(\theta+\epsilon-\frac{\pi}{4}) + \gamma_3\mathcal{R}(\theta+\epsilon+\pi).
\label{eq:ourmixture}
\end{align}
The coefficients $\gamma_1, \gamma_2, \gamma_3$ take the following form:
\begin{align} 
\gamma_1(\epsilon) &= \frac{1}{2} \left(\cos (\epsilon)-\cot \left(\frac{\pi }{8}\right) \sin (\epsilon)+1\right) \nonumber \\
\gamma_2(\epsilon) &= \sqrt{2}\sin(\epsilon) \label{eq:ourgammas} \\
\gamma_3(\epsilon) &=
\sec \left(\frac{\pi }{8}\right) \sin \left(\frac{\epsilon}{2}\right) \sin \left(\frac{1}{8} (4 \epsilon-\pi )\right) \nonumber,
\end{align}
where $\epsilon$ is a positive real number. The decomposition for negative $\epsilon$ is analogous (see Appendix \ref{ap:opt_sol}).

Note that one of the $\{\gamma_i\}$ must be negative, meaning that the decomposition in Eq.~\ref{eq:ourmixture} is not a probability distribution. This motivates us to build the unbiased estimator for $\mathcal{R}(\theta)$ next.

Denoting $\widetilde{\mathcal{R}}(\theta) \equiv \mathcal{R}_1(\theta)$, $\widetilde{\mathcal{R}}(\theta -\frac{\pi}{4}) \equiv \mathcal{R}_2(\theta)$, and $\widetilde{\mathcal{R}}(\theta + \pi) \equiv \mathcal{R}_3(\theta)$, and using Eq.~\ref{eq:ourmixture} and Eq.~\ref{eq:ourgammas}, the probability of sampling $\mathcal{R}_i$ is $|\gamma_i(\epsilon)| / \|\gamma(\epsilon)\|_1$, where  $\gamma(\epsilon )\equiv$ 
$\begin{bmatrix} \gamma_1(\epsilon) & \gamma_2(\epsilon) & \gamma_3(\epsilon) \end{bmatrix}^T$.
This allows us to construct an estimator for the noiseless channel, $\mathcal{R}(\theta)$ as:
\begin{equation}
\hat{\mathcal{R}}(\theta) = \|\gamma(\epsilon)\|_1 \text{sign}[{\gamma_i(\epsilon)}]\mathcal{R}_i(\theta),
\label{eq:gateestimator}
\end{equation}
where $i \in \left\{1,2,3\right\}$ and, as desired, $\mathbb{E}[\hat{\mathcal{R}}(\theta)] = \mathcal{R}(\theta)$ (see Appendix\ref{ap:est} for proof).

\subsection{Mixture of twirled noisy channels}\label{subsec:mix_twirl}
Given a single-qubit noisy gate $\bar{U}$,
\begin{equation}\label{eq:noisy_CT_gates}
    \bar{U}= R(\epsilon') R(\theta) =
    e^{-i\frac{\epsilon_z}{2}\mathbb{Z}}
    e^{-i\frac{\epsilon_y}{2}\mathbb{Y}}
    e^{-i\frac{\epsilon_x}{2}\mathbb{X}}
    e^{-i\frac{\theta}{2}\mathbb{P}},
\end{equation}
where $\mathbb{P}$ is a Pauli operator, $\{\theta, \epsilon_x, \epsilon_y, \epsilon_z, \epsilon'\}$ are real numbers satisfying $\epsilon_1^2 + \epsilon_2^2 + \epsilon_3^2 = (\epsilon')^2$.
We define the twirled noisy channel as follows:
\begin{equation}
    \mathcal{T}_{\bar{U}}(.) = \frac{1}{|\mathcal{G}|} \sum_{\sigma\in\mathcal{G}} \sigma \bar{U} \sigma^{\dagger} (.) \sigma^{\dagger} \bar{U}^{\dagger} \sigma,
\end{equation}
where $\mathcal{G}$ is the set of Pauli operators which commute with $\mathbb{P}$ and $|\mathcal{G}|$ is its size.

For the case when $\mathbb{P}=\mathbb{Z}$, $\mathcal{G} = \{\mathbb{I}, \mathbb{Z}\}$, so
\begin{equation}
    \mathcal{T}_{\bar{U}}(.) = \frac{1}{2} \sum_{\sigma\in\{\mathbb{I}, \mathbb{Z}\}} \sigma \bar U' \sigma^{\dagger} \widetilde{U}' (.) \widetilde{U}'^{\dagger} \sigma^{\dagger} \bar{U}'^{\dagger} \sigma,
\end{equation}
where $\widetilde{U}^{'}=R_z(\theta+\epsilon_z)$ and $\bar{U}'=e^{-i\frac{\epsilon_y}{2}\mathbb{Y}}e^{-i\frac{\epsilon_x}{2}\mathbb{X}}$ are unitary gates.

For small values of $\epsilon_x$ and $\epsilon_y$, twirling mitigates $\bar U'$ the following approximation holds:
\begin{equation}
    \mathcal{T}_{\bar{U}}(.) \approx \widetilde{\mathcal{U}}'(.)= {R}_z(\theta+\epsilon_z) (.) {R}^{\dagger}_z(\theta+\epsilon_z).
\end{equation}
Thus, we design a new estimator for the ideal channel, $\mathcal{R}_z(\theta)$, using the twirled noisy channel as follows:
\begin{equation}\label{eq:CT_gateestimator}
    \hat{\mathcal{R}}_z(\theta) = \|\gamma(\epsilon_z)\|_1 \text{sign}[{\gamma_i}(\epsilon_z)] \bar{\mathcal{R}}_{z_{i}}^{\sigma_j}(\theta),
\end{equation}
where $\sigma_j \in  \{\mathbb{I}, \mathbb{Z}\}$, $z_i \in \{z_1, z_2, z_3\}$, and the channel defined as:
\begin{align}
    \bar{\mathcal{R}}_{z_{i}}^{\sigma_j}(\theta) (.) &= \sigma_j \bar{R}_{z_{i}}(\theta) (.) \bar{R}_{z_{i}}^{\dagger}(\theta) \sigma_j^{\dagger}, \label{eq:CT_noisychannel} \\
    \text{with }
    \bar{R}_{z_{1}}(\theta)
    &=R_z(\theta+\epsilon_z)\bar{U}', \nonumber \\
    \bar{R}_{z_{2}}(\theta)
    &=R_z(\theta+\epsilon_z - \frac{\pi}{4})\bar{U}', \nonumber \\
    \text{and }
    \bar{R}_{z_{3}}(\theta) 
    &=R_z(\theta+\epsilon_z+\pi)\bar{U}'\nonumber \\
    \bar{U}' 
    &= 
    e^{-i\frac{\epsilon_y}{2}\mathbb{Y}}
    e^{-i\frac{\epsilon_x}{2}\mathbb{X}}. \nonumber
\end{align}
If the probability of choosing the channel $\bar{\mathcal{R}}_{z_{i}}^{\sigma_j}(\theta)$ is fixed to be $|\gamma_i(\epsilon_z)| / 2 \|\gamma(\epsilon_z)\|_1$, where $\gamma(\epsilon_z) =\begin{bmatrix} \gamma_1(\epsilon_z) & \gamma_2(\epsilon_z) & \gamma_3(\epsilon_z) \end{bmatrix}^T$, it can be seen that $\mathbb{E}[\hat{\mathcal{R}}_z(\theta)] \approx \mathcal{R}_z(\theta)$ (see Appendix\ref{ap:est} for proof). 

\subsection{Mixture of noisy circuits}\label{sec:mix_noisy_circuits}
Given a quantum circuit $U_{c} = \prod_j U_j$ which contains $\nu$ gates of the form $U_{j} = R_j(\theta_j) = e^{-i\frac{\theta_j}{2}\mathbb{P}_j}$, where $\theta_j \in [0, 2\pi)$ and $\mathbb{P}_j$ is a Pauli-string,
we can construct an estimator for the corresponding quantum channel $\mathcal{U}_c$ using the estimators in either Eq.~\ref{eq:gateestimator} or Eq.~\ref{eq:CT_gateestimator}.

Using the gate estimator in Eq.~\ref{eq:gateestimator}, the estimator for $\mathcal{U}_c$ can written as follows:
\begin{align} 
\hat{\mathcal{U}}_c &= \prod_{j=1}^{\nu} \hat{\mathcal{R}}^{(j)}(\theta_j) \nonumber\\
&= \prod_{j=1}^{\nu} \|\gamma^{(j)}(\epsilon_j)\|_1 \text{sign}[{\gamma_{l_j}^{(j)}{(\epsilon_j)}}]\mathcal{R}^{(j)}_{l_j}(\theta_j) \nonumber\\
&= \prod_{j=1}^{\nu} \|\gamma^{(j)}(\epsilon_j)\|_1 \text{sign}[ \prod_{j=1}^{\nu} {\gamma_{l_j}^{(j)}{(\epsilon_j)}}] \prod_{j=1}^{\nu}\mathcal{R}^{(j)}_{l_j}(\theta_j) \nonumber \\
&= \Gamma_c \: \text{sign}[\Gamma_L] \mathcal{U}_L, 
\label{eq:circuit_estimator}
\end{align}
where $L = \{l_1,...,l_{\nu}\}$ is a sequence of indices for each sampled gate with $l_i \in \left\{1,2,3\right\}$ and $\Gamma_c = \prod_{j=1}^{\nu} \|\gamma^{(j)}(\epsilon_j)\|_1$.
The probability of sampling a noisy circuit $\mathcal{U}_L$ is hence $p_L = |\Gamma_L|/\Gamma_c$. In Appendix\ref{ap:est}, we prove that $\mathbb{E}[\hat{\mathcal{U}}_c] = {\mathcal{U}}_c$.
All circuits $\mathcal{U}_L$ consist of gates that undergo a coherent error of the form shown in Eq.~\ref{eq:cohe_err}.

Similarly, a circuit estimator can be constructed using gates of the form given in Eq.~\ref{eq:noisy_CT_gates}, by replacing the gate estimators above with those defined in Eq.~\ref{eq:CT_gateestimator}.

\subsection{Observable estimation}
Given access to a noisy circuit, $\mathcal{U}'_c$, composed of noisy channels as in Eq.~\ref{eq:mixture} or Eq.~\ref{eq:CT_noisychannel}, we can estimate the expectation value of a Hermitian observable, $\langle \mathcal{O} \rangle_{\mathcal{U}_c} = \text{Tr}[\mathcal{O}\mathcal{U}_c\ketbra{0}{0}]$ corresponding to the ideal circuit $\mathcal{U}_c$.
For a complete set of positive measurement operators $\{E_b\}$, satisfying $\mathcal{O} = \sum_b\text{Tr}[\mathcal{O}E_b] E_b$, we obtain:
\begin{align}
    \braket{\mathcal{O}}_{\mathcal{U}_c} &=  \text{Tr}[\mathcal{O}\mathcal{U}_c\ketbra{0}{0}] \nonumber\\
    &= \text{Tr}[ (\sum_b \text{Tr}[\mathcal{O}E_b] E_b) \mathcal{U}_c \ketbra{0}{0}] \nonumber \\
    &= \sum_b \text{Tr}[\mathcal{O}E_b] \text{Tr}[E_b \mathcal{U}_c  \ketbra{0}{0}] \nonumber \\
    &= \sum_b p_b \text{Tr}[\mathcal{O}E_b],
\end{align}
where $p_b=\text{Tr}[ E_b \mathcal{U}_c\ketbra{0}{0}]$ is the probability of observing the outcome $\ket{b}$.

Now, we can construct an estimator for the expectation value as:
\begin{equation}
    \hat{o} =  \Gamma_c \: \text{sign}[\Gamma_L] \text{Tr}[\mathcal{O}E_b],
\label{eq:exp_estimator}
\end{equation}
where $E_b$ is the measurement operator, and $\Gamma_L$, $\Gamma_c$ are defined in Eq.~\ref{eq:circuit_estimator}.
Using the estimator in Eq.~\ref{eq:exp_estimator}, we can calculate the ideal expectation value (see Appendix\ref{ap:est} for more details), albeit at the cost of increasing sampling cost. 
We summarize this trade-off in a statement that follows. 

\begin{statement}\label{statement}
Using the probabilistic mixture defined in Eq.~\ref{eq:ourmixture}, one can calculate the expectation value of a Hermitian operator $\mathcal{O}$ given by $\braket{{\mathcal{O}}} = \text{Tr}[\mathcal{O}\mathcal{U}_c\ketbra{0}{0}]$, when only provided access to noisy circuits $\mathcal{U}_L$ containing $\nu$ unitaries subject to a unitary over-rotation through an angle $\epsilon$. The  number of circuit runs $S$ required to estimate $\braket{\mathcal{O}}$ up to error $\delta$ is bounded by 
\begin{equation}
S \leq \frac{e^{0.83 |\epsilon|\nu}\|\mathcal{O}\|_{\infty}^2}{ \delta^2}.
\label{eq:shotsscaling}
\end{equation}
\end{statement}

The proof of Statement~\ref{statement} is outlined in Appendix\ref{ap:overhead}. 
The limitations of our method become apparent in Eq.~\ref{eq:shotsscaling}: when the number of noisy rotations $\nu$ or the error $\epsilon$ is too large, the resulting high variance can degrade the accuracy of the expectation value estimate. 

However, the method remains applicable to circuits of moderate size and error strength. Thus, in Sec.~\ref{sec:results} we consider cases where $\nu$ and $|\epsilon|$ are on the orders of $10^3$ and $10^{-3}$, respectively.

\section{Numerical Simulations}\label{sec:results}

In this section, we present results from different numerical simulations to show empirical evidence supporting the method proposed in this paper.
We carried out all the simulations using the \textsc{Qiskit}~\cite{qiskit2024} package.

As outlined in the previous section, we construct unbiased estimators using prior knowledge of the error magnitude associated with each gate. 
Consequently, the proposed technique is applicable in scenarios where the error channel is known a priori. 
A natural application of this approach is the reduction of error in expectation value estimation within a fault-tolerant framework. 
In this setting, continuous rotations are approximated using a universal, finite gate set resulting in a constant error at each gate (as described in Eq.~\ref{eq:noisy_CT_gates}), with the corresponding error parameters known exactly.
We further demonstrate the applicability of our method in the near-term regime, wherein each physical gate is subject to coherent error. 
These errors typically arise from inaccuracies in classical control of the quantum hardware and, in principle, can be mitigated when the error channel is known.
Accordingly, the fault-tolerant setting constitutes the primary intended application of our method.

To evaluate the performance of the method, we use the time dynamics simulation of the one-dimensional Ising-like Hamiltonian with periodic boundary conditions:
\begin{equation}
H = h\sum_{i=1}^{N}\mathbb{Y}_i + J\sum_{i=1}^{N}\mathbb{X}_i\mathbb{X}_{i+1},
\label{eq:hamiltonian}
\end{equation}
where qubit $N+1$ is identified with qubit 1.

For all the numerical experiments, we fix $h=J=1$.
The circuit for the simulation, $\mathcal{C}$, is constructed using first-order Trotterization with $\mathcal{L}$ Trotter steps:
\begin{align}
\mathcal{C} &= \left\{e^{-iH\frac{T}{\mathcal{L}}}\right\}^{\mathcal{L}}\nonumber \\
&= \left\{ \prod_{i=1}^{N} e^{-i{\mathbb{X}_i\mathbb{X}_{i+1}}\frac{T}{\mathcal{L}}} \prod_{i=1}^{N} e^{-i\mathbb{Y}_i\frac{T}{\mathcal{L}}}\right\}^{\mathcal{L}} \nonumber \\
&= \left\{\prod_{i=1}^{N} R_{x_{i} x_{i+1}}
\left( \frac{2T}{\mathcal{L}} \right)
\prod_{i=1}^{N} R_{y_i}
\left(\frac{2T}{\mathcal{L}}\right) \right\}^{\mathcal{L}},
\label{eq:circuit}
\end{align}
where $T$ is the evolution time.
It can be seen that the total number of  parametrized  gates is $\nu = 2N\mathcal{L}$. After evolving $\ket{0}^{\otimes N}$ through the circuit in Eq.~\ref{eq:circuit}, we measure $\braket{\mathcal{O}}=\braket{\mathbb{Z}^{\otimes N}}$.

In order to apply our method to estimate an expectation value, one needs to specify two parameters: the total number of circuit runs, $S$, and the number of shots per sampled circuit instance, $s$. 
So, the total number of circuit instances required for our method is $N_c \equiv S/s$. 
In contrast, only one circuit is needed to evaluate the expectation value in noiseless and noisy cases. We set $s$ to $100$ in all simulations. 
This value was chosen based on the simulations in Appendix\ref{ap:num_circuits}.

In the following sections, we provide detailed descriptions of the simulations conducted and present a comprehensive analysis of the results.
For circuit constructions and additional details on the visualization of results, we refer the reader to Appendices \ref{ap:circuitry} and \ref{ap:visualizing}, respectively.

\subsection{Approximation error}\label{subsec:approx_error}
An arbitrary single-qubit rotation $U \in \mathrm{SU}(2)$ can be approximated using a universal gate set $G$ to accuracy $\zeta$ by a finite sequence of gates from $G$ of length $\mathcal{O}(\log^c(1/\zeta))$, where the constant $c$ depends on the synthesis strategy and ranges from $1$~\cite{ross2014optimal} to $3.97$~\cite{kitaev1997quantum, dawson2005solovay}.
As discussed in Sec.~\ref{subsec:aus}, the approximate unitary can be written as $V = U'U$ with
\begin{equation}
\|\mathbb{I} - U'\| < \zeta,
\label{eq:operatornorm}
\end{equation}
where $\zeta$ is a positive number chosen to achieve desired accuracy and $U'$ is the approximation error.
In the context of fault-tolerant quantum computation, this systematic error accumulates and can only be mitigated by increasing the resource overhead.
Our method enables the efficient and accurate estimation of observables with lower overhead.
In what follows, we demonstrate the utility of our method for this task.

\subsubsection{Modeled decomposition error}

In this section, we model the approximation error $U'$ as a unitary rotation of the form as $U' = e^{-i\epsilon R'}$ (Eq.~\ref{eq:approx_unit_err}), where $\epsilon$ is a real number. 
We can then calculate the expression for the error as follows:

\begin{align}
\|\mathbb{I} - U'\| &= \|\mathbb{I} - e^{-i\epsilon \boldsymbol{\eta} \cdot \boldsymbol{\sigma}}\|
\nonumber \\
&= 2\sin(|\epsilon|/4) \approx |\epsilon|/2 \nonumber\\
&< \zeta,
\label{eq:operatornormformula}
\end{align}
where $\boldsymbol{\eta} = \begin{bmatrix} \eta_x & \eta_y & \eta_z \end{bmatrix}^T$ obeys $\eta_{x}^{2} + \eta_{y}^{2} + \eta_{z}^{2} = 1$ and $\boldsymbol{\sigma} = \begin{bmatrix} \mathbb{X} & \mathbb{Y} & \mathbb{Z} \end{bmatrix}^T$.
Assuming $\epsilon$ is small, the approximation unitary can also be written as:
\begin{equation}
    U' \approx R_{z}(\epsilon_{z})R_{y}(\epsilon_{y})R_{x}(\epsilon_{x}),
\label{eq:unstructured_error}
\end{equation}
where $\epsilon_{x}=\eta_x \cdot\epsilon, \epsilon_{y}=\eta_y \cdot\epsilon$ and $\epsilon_{z}=\eta_z \cdot\epsilon$.

To study the behavior of $\braket{\mathcal{O}}$ under an approximation error, we first decompose the circuit in Eq.~\ref{eq:circuit} using Clifford gates and $R_z$ rotation. 
Next, we find an approximate decomposition of the $R_z$ rotation gate as $\bar{R}_z = U'R_z$.
Using these approximate gates the noisy circuit corresponding to the ideal circuit in Eq.~\ref{eq:circuit} can be written as:
\begin{align}
  \bar{\mathcal{C}} = \Bigg\{ &\prod_{i=1}^{N} C^{(2)}_{i,i+1} \bar{R}_{z_i}
  \left(\frac{2T}{\mathcal{L}}\right) C^{(2)\dagger}_{i,i+1} \nonumber \\
  &\times \prod_{i=1}^{N} C^{(1)}_i\bar{R}_{{z}_i}
  \left(\frac{2T}{\mathcal{L}}\right)
  C^{(1)\dagger}_i \Bigg\}^{\mathcal{L}},
\label{eq:noisy_app_circuit}
\end{align}
where $C^{(1)}_i = \mathbb{H}_i \mathbb{S}^{\dagger}_i$ and $C^{(2)}_{i,i+1} = \mathbb{H}_i \mathbb{H}_{i+1} \text{CNOT}(i,i+1)$ are Clifford transformations, in which $\mathbb{S}$, $\mathbb{H}$, and CNOT denote the phase, Hadamard, and controlled-NOT gates, respectively.

We choose the values of $\epsilon_{x}, \epsilon_{y}$ and $\epsilon_{z}$ such that they satisfy the following relation:
\begin{equation}
    \sqrt{\epsilon_{x}^2 + \epsilon_{y}^2 + \epsilon_{z}^2} = |\epsilon| < 2\zeta.
\label{eq:epsilon_zeta}
\end{equation}
This relation ensures that the approximation error $\zeta$ is bounded by the desired accuracy. The values for these undesired over-rotations are known exactly for a given decomposition of noiseless gate. They can be suppressed using Pauli twirling and our proposed method as discussed in Sec.~\ref{subsec:mix_twirl} and Appendix\ref{ap:chan_decomp}.  Importantly, both strategies operate entirely within the Clifford+$T$ gate set.

In our case, the noiseless operation is $R_z(\theta)$, so the commuting Pauli operators are simply $\mathbb{I}$ and $\mathbb{Z}$, and twirling can suppress undesired rotations in the $X$- and $Y$- directions. 
Meanwhile, the residual over-rotation in $Z$-direction can be removed by probabilistically inserting $R_z(\pi) = -i\mathbb{Z}$ and $R_z(-\pi/{4}) = e^{-i\pi/8 }T^\dagger$ (for positive $\epsilon_z$) or $R_z(\pi/{4}) = e^{i\pi/8}T$ (for negative $\epsilon_z$) according to Eq.~\ref{eq:ourmixture}. 
This process recovers the original channel $\mathcal{R} _z(\theta)$ approximately as described in Appendix\ref{ap:est}.

We now perform simulations to calculate the expectation value of an observable $\braket{\mathbb{Z}^{\otimes N}}$.
For this numerical experiment, we fix $N=15$ in the Hamiltonian (Eq.~\ref{eq:hamiltonian}) and simulate the time evolution using $\mathcal{L}=70$ Trotter steps. 
We fixed the values of $\epsilon$ to be 0.001 and applied the same noise $U' = R_z(\epsilon_{z}) R_y(\epsilon_{y}) R_x(\epsilon_{x})$ to every $R_z(2T/\mathcal{L})$ gate, so that the values of $\epsilon_{x}, \epsilon_{y}$ and $\epsilon_{z}$ satisfy $\epsilon_x^2 + \epsilon_y^2 + \epsilon_z^2 = \epsilon^2$. This noise mimics the finite-gate decomposition of a continuous rotation. The circuit contains $\nu = 2100$ rotation gates, and a total of $S=10^5$ shots were used to calculate $\braket{\mathcal{O}}$. This calculation was performed using the exact and noisy circuits, as well as our method, both alone and in combination with Pauli twirling. The number of noisy circuit instances used for our method was $10^3$. See Appendix\ref{ap:circuitry} for more details on circuit construction.

\begin{figure}[htbp!]
    \centering
    \includegraphics[width=0.49\textwidth]{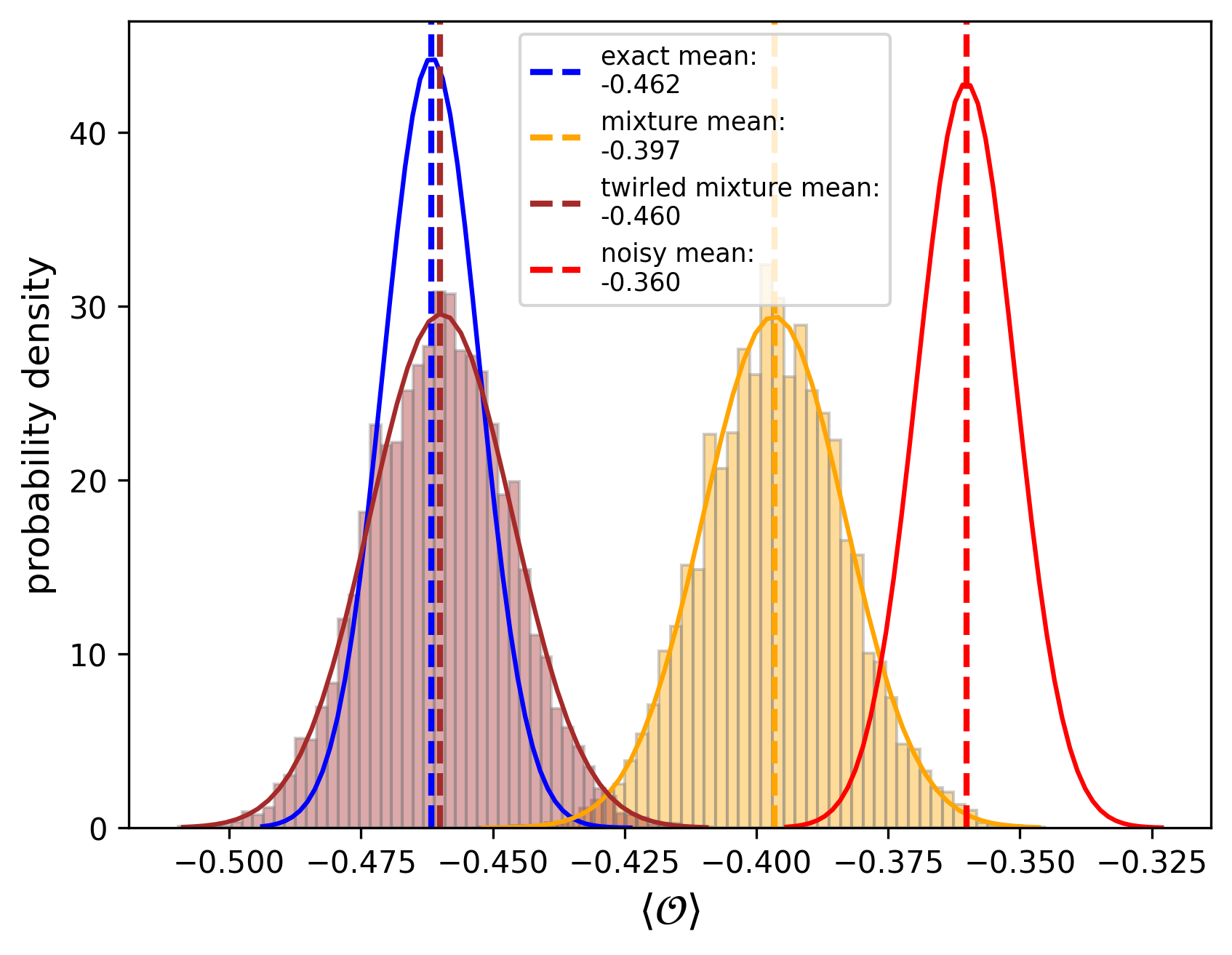}
    \caption{
    Estimating the expectation value $\braket{\mathcal{O}}$ in the presence of unitary errors along the $X$-, $Y$-, and $Z$-axes. Unitary error, modeled by Eq.~\ref{eq:unstructured_error} and applied at every  parametrized  gate in the circuit (Eq.~\ref{eq:noisy_app_circuit}), causes significant deviation from the exact value (dashed blue line). The noisy expectation value (dashed red line) demonstrates this deviation. By employing our method, the estimated expectation value moves closer to the exact one. Further improvement is observed when combining the proposed strategy with Pauli twirling, as indicated by the brown dashed line closely aligning with the blue dashed line. However, the variance of the expectation value distribution increases when probabilistic mixtures are applied.
    }
\label{fig:twirlingerrormitigation}
\end{figure}

The results of this numerical experiment are shown in Figure~\ref{fig:twirlingerrormitigation}.
We observe that the approximation error due to decomposition leads to a shift in the distribution.
Using the estimators in Eq.~\ref{eq:exp_estimator}, we observe that the distribution shifts towards the ideal distribution, however it is still not close to the ideal one.
The results with estimators constructed using the combination of our approach with Pauli twirling (Eq.~\ref{eq:CT_gateestimator}) closely align with the ideal distribution and accurately predict the expectation value.
As expected, we observe that the variance of the distribution increases when compared to the ideal distribution.

Next, we perform simulations to track the evolution of the expectation value $\braket{\mathcal{O}}$ after each Trotter step.
This allows us to analyze the effect of the approximation error as the number of gates increases. 
For this, we use a Hamiltonian (Eq.~\ref{eq:hamiltonian}) on $N=12$ qubits and simulate the evolution up to $\mathcal{L}=30$ Trotter steps. 
For each simulation, we chose new noise parameters $\epsilon_x, \epsilon_y, \epsilon_z$ for our gate $R_z(2T/\mathcal{L})$, each time satisfying $\epsilon_x^{2}+\epsilon_y^{2}+\epsilon_z^{2}=\epsilon^2$, where the value of $\epsilon$ is $0.02$. 
This can be viewed as using different finite-gate-set approximations with decomposition error $\zeta = \epsilon/2$. 

\begin{figure}[htbp!]
    \centering
    \includegraphics[width=0.49\textwidth]{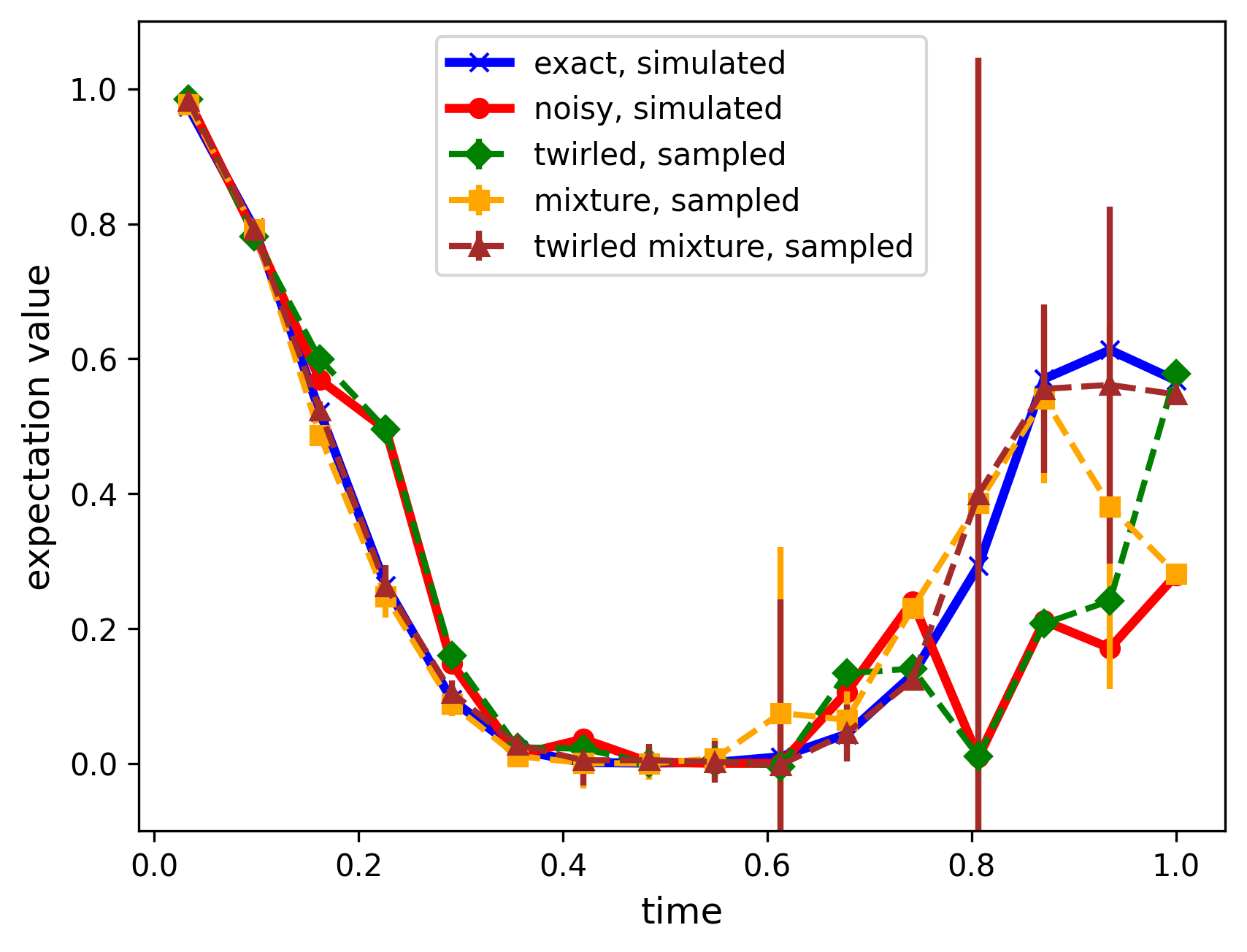}
    \caption{
    Time evolution of $\braket{\mathbb{Z}^{\otimes N}}$ in the presence of larger unitary errors ($\epsilon = \sqrt{\epsilon_x^2 + \epsilon_y^2 + \epsilon_z^2} = 0.02$) applied at every  parametrized  rotation in the circuit (Eq.\ref{eq:noisy_app_circuit}). 
    The unstructured error (Eq.\ref{eq:unstructured_error}) models over-rotations due to decomposition with universal finite gate sets, causing noisy expectation values (red curve) to fluctuate around the exact values (blue curve).
    Combining our method with Pauli twirling significantly improves the estimation of the expectation value (brown curve), outperforming the use of either technique alone (orange and green curves). The resulting estimate closely tracks the ideal evolution of $\braket{\mathbb{Z}^{\otimes N}}$.
    Times 0.8 and 1 are the cases where $\epsilon_z$ contributes the most and the least to the overall error, respectively.
    }
\label{fig:twirledmixturestepbystep}
\end{figure}

The results from this simulation are presented in Figure~\ref{fig:twirledmixturestepbystep}. Exact and noisy expectation values were calculated precisely (with infinite shots) at each time. To estimate $\braket{\mathcal{O}}$ using our method (with and without twirling), we utilized $10^5$ circuit runs, sampling the total of $10^3$ circuit instances for each time step.
It can be seen that the approximation error causes the expectation value to fluctuate about the ideal one. 
However, with the estimators constructed through the combination of our approach and Pauli twirling (Eq.\ref{eq:CT_gateestimator}), the expectation value calculated using noisy circuits closely follows the ideal curve. 
Moreover, when using either Pauli twirling alone or the estimators in Eq.\ref{eq:exp_estimator}, the expectation value still fluctuates about the ideal quantity, showing accuracy in some instances while exhibiting deviations in others.

If we examine two points on the curve, at times 0.8 and 1 in Figure~\ref{fig:twirledmixturestepbystep}, distinct behaviors can be observed.
At time 0.8, $\epsilon_z$ is the dominant error contribution. 
Under this condition, twirling offers no improvement since $\mathbb{I}$ and $\mathbb{Z}$ commute through the noisy $R_z(\theta+\epsilon_z)$.
Consequently, the red and green data points in Figure~\ref{fig:twirledmixturestepbystep} correspond to the same expectation value close to $0$. 
In this scenario, the approximation error is suppressed solely by our method, leading to high variance, with the brown and orange data points overlapping in Figure~\ref{fig:twirledmixturestepbystep}.
At time 1, the opposite condition occurs: $\epsilon_z \approx 0$. 
Here, our method does not contribute to suppressing the error, as indicated by red and orange data points in Figure~\ref{fig:twirledmixturestepbystep} predicting the same expectation value of 0.28. 
Instead, twirling plays the primary role in recovering the exact expectation value, as demonstrated by the alignment of the brown and green data points in Figure~\ref{fig:twirledmixturestepbystep}.

\subsubsection{Clifford+$T$ gate set decomposition error}

In this section, we use the algorithm from Ref.\cite{ross2014optimal} to approximate $R_z$ gates with varying accuracy, $\zeta > \|R_z(\theta) - \bar R_z(\theta)\|$.
We then carry out numerical simulations, fixing $N=12$ in the Hamiltonian (Eq.~\ref{eq:hamiltonian}) and use $\mathcal{L}=24$ Trotter steps for an evolution time $T=0.8$, resulting in a total of $\nu = 576$ rotations, all with $\theta = 0.067$. 
The noisy unitary $\bar R_z(\theta) = U'R_z(\theta)$ is implemented as a single-qubit Clifford+$T$ sequence.
To use our method, we model the unitary error as $U' = R_z(\epsilon_z)R_y(\epsilon_y)R_x(\epsilon_x)$ (Eq.~\ref{eq:unstructured_error}). 
Since both $R_z(\theta)$ and $\bar R_z(\theta)$ are known exactly, we compute the angles $\epsilon_z$, $\epsilon_y$, and $\epsilon_x$ directly and use them to evaluate the coefficients $\{\gamma_1(\epsilon_z), \gamma_2(\epsilon_z), \gamma_3(\epsilon_z)\}$ (Eq.~\ref{eq:ourgammas}).
We then apply our method, combined with Pauli twirling as in the previous section, to mitigate the approximation error.

\begin{figure}[htbp!]
    \centering
    \includegraphics[width=0.49\textwidth]{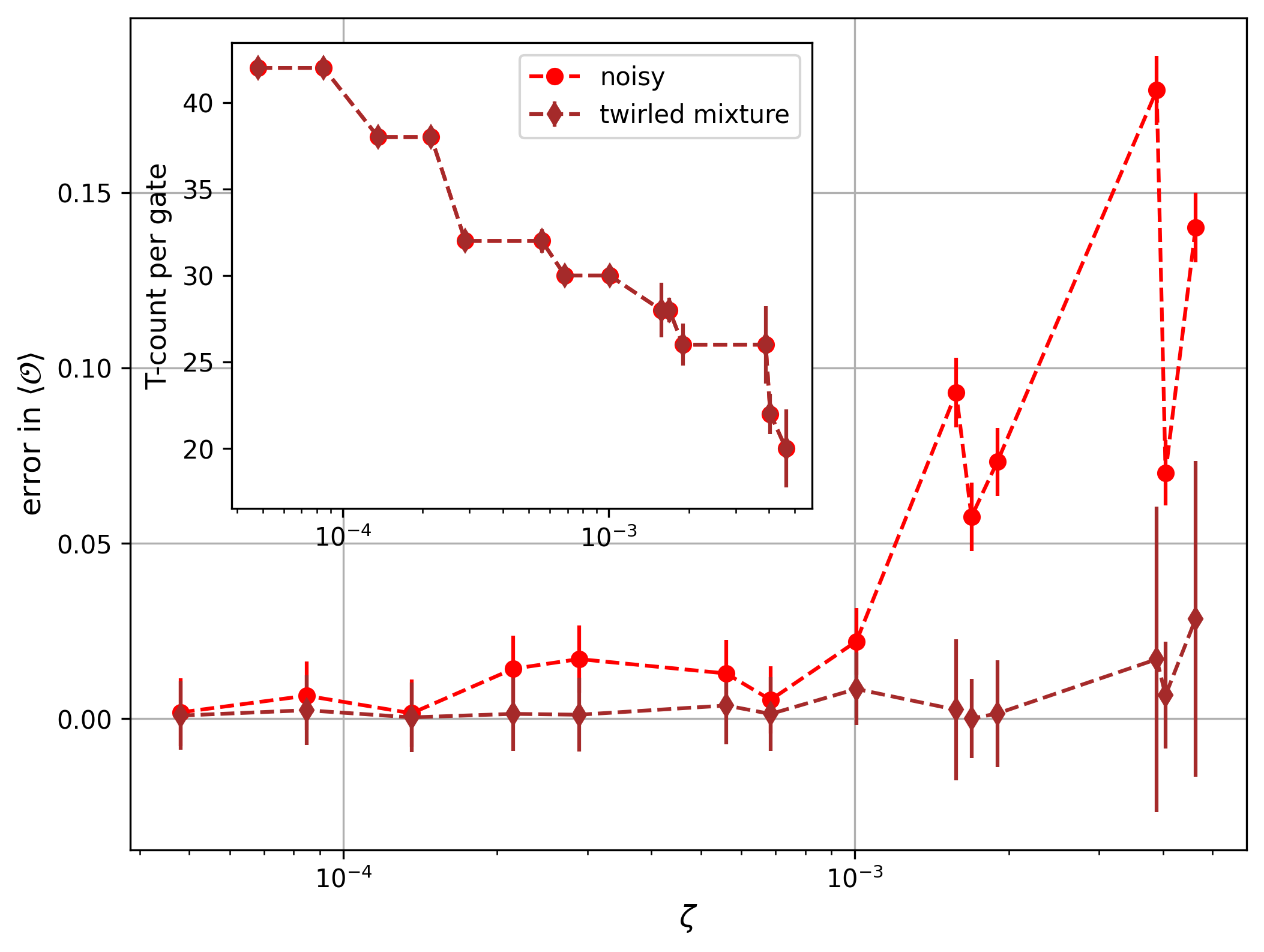}
    \caption{Absolute error in the expectation value, $|\braket{\mathcal{O}} - \braket{\mathcal{O}}_{\text{noisy}}|$, with different Clifford+$T$ decomposition accuracy, $\zeta$.  The main plot compares our method, combined with Pauli twirling, against unmitigated expectation value estimates. A significant improvement is observed when using probabilistic mixtures, though at the cost of increased variance, as $\epsilon_z$ grows with $\zeta$ on average (Eq.~\ref{eq:epsilon_zeta}). 
    The inset plot shows the number of $T$ gates per $R_z(\theta)$ in the Clifford+$T$ decomposition, both with and without our method. Since the $T$-gate overhead introduced by our approach is small (Eq.~\ref{eq:t_overhead}), the red and brown data points largely overlap. The significant improvement in the expectation value estimate is observed for $\zeta > 10^{-3}$.
    }
\label{fig:sk_errormitigation}
\end{figure}

The results from this numerical experiment are presented in Figure~\ref{fig:sk_errormitigation}. 
A total of $S = 10^5$ circuit runs were used to estimate the observable $\braket{\mathcal{O}} = \braket{\mathbb{Z}^{\otimes N}}$ at each data point. 
In our method, we used $s = 100$ shots per sampled circuit instance.
The significant improvement achieved by the proposed technique is evident in the main plot of Figure~\ref{fig:sk_errormitigation}. 
The inset plot displays the number of $T$ gates required in the Clifford+$T$ decomposition of $R_z(\theta)$. 
As shown, the $T$-gate overhead per  parametrized  gate, denoted $\Omega_T$, introduced by our method is negligible. 
It can be analytically calculated as:
\begin{align}
    \Omega_T(\epsilon) &= \frac{|\gamma_2(\epsilon)|}{\|\gamma(\epsilon)\|_1} = \sqrt{2} \cos \left(\frac{\pi }{8}\right) \sin (|\epsilon|) \sec \left(|\epsilon|-\frac{\pi }{8}\right) \nonumber \\
    &\approx \sqrt{2}\, |\epsilon|,
\label{eq:t_overhead}
\end{align}
where the expressions for $\{\gamma_i\}$ are given in Eq.~\ref{eq:ourgammas} and, in our case, $|\epsilon| = |\epsilon_z|$ was at most $3 \times 10^{-3}$. Thus, each circuit instance sampled using our method required fewer than $\Omega_T^{\text{max}}\nu < 3$ additional $T$ rotations, on top of the $T$-count from the Clifford+$T$ decomposition.

Figure~\ref{fig:sk_errormitigation} illustrates that combining our strategy with Pauli twirling preserves the accuracy of expectation value estimates across a range of decomposition errors. While improvements are noticeable for $\zeta > 10^{-4}$, they become especially significant beyond $\zeta > 10^{-3}$, where the $T$-count drops below 30. In this regime, the absolute error in the unmitigated (noisy) estimate increases by an order of magnitude, whereas the mitigated result maintains a reasonable level of accuracy, even when the number of $T$ gates is reduced from 30 to 20. These results suggest our method enables accurate observable estimation with low $T$-overhead, making it particularly useful in the early fault-tolerant quantum era.

\subsection{Coherent over-rotation}\label{subsec:coherent_or}
In this section, we study the case in which every gate in the simulation circuit, $\mathcal{C}$ (Eq.~\ref{eq:circuit}), is subject to a constant over-rotation, so the noisy circuit can be written as
\begin{equation}
  \widetilde{\mathcal{C}} = \left\{ \prod_{i=1}^{N} R_{x_{i} x_{i+1}}\left(\frac{2T}{\mathcal{L}}+\epsilon_{i, i+1}\right) \prod_{i=1}^{N} R_{y_i}\left(\frac{2T}{\mathcal{L}}+\epsilon_i\right)\right\}^{\mathcal{L}},
\label{eq:noisy_circuit}
\end{equation}
where $\{\epsilon_{i,j}\}$ and $\{\epsilon_i\}$ are real numbers.
We consider two different cases: one where the value of $\epsilon$ is known exactly, and another where only the mean value of the distribution for $\epsilon$ is known. The latter scenario is presented in Appendix\ref{ap:add_num_exp}.

Here, we assume that over-rotation $\epsilon=0.001$ is known precisely and is constant for all gates in the circuit (Eq.~\ref{eq:noisy_circuit}).
We fix $N=15$ in the Hamiltonian (Eq.~\ref{eq:hamiltonian}) and simulate the time evolution using $\mathcal{L}=70$ trotter steps utilizing $S=3\times 10^6$ shots for calculating the expectation value.
We begin with estimating $\braket{\mathcal{O}}$ using exact (noiseless) circuit.
We then estimate $\braket{\mathcal{O}}$ using the noisy circuit in Eq.~\ref{eq:noisy_circuit}.
Finally, we apply the estimator in Eq.~\ref{eq:exp_estimator} to evaluate $\braket{\mathcal{O}}$, sampling $N_c = 3 \times 10^4$ circuit instances according to the method outlined in Sec.~\ref{sec:mix_noisy_circuits}.

\begin{figure}[htbp!]
    \centering
    \includegraphics[width=0.50\textwidth]{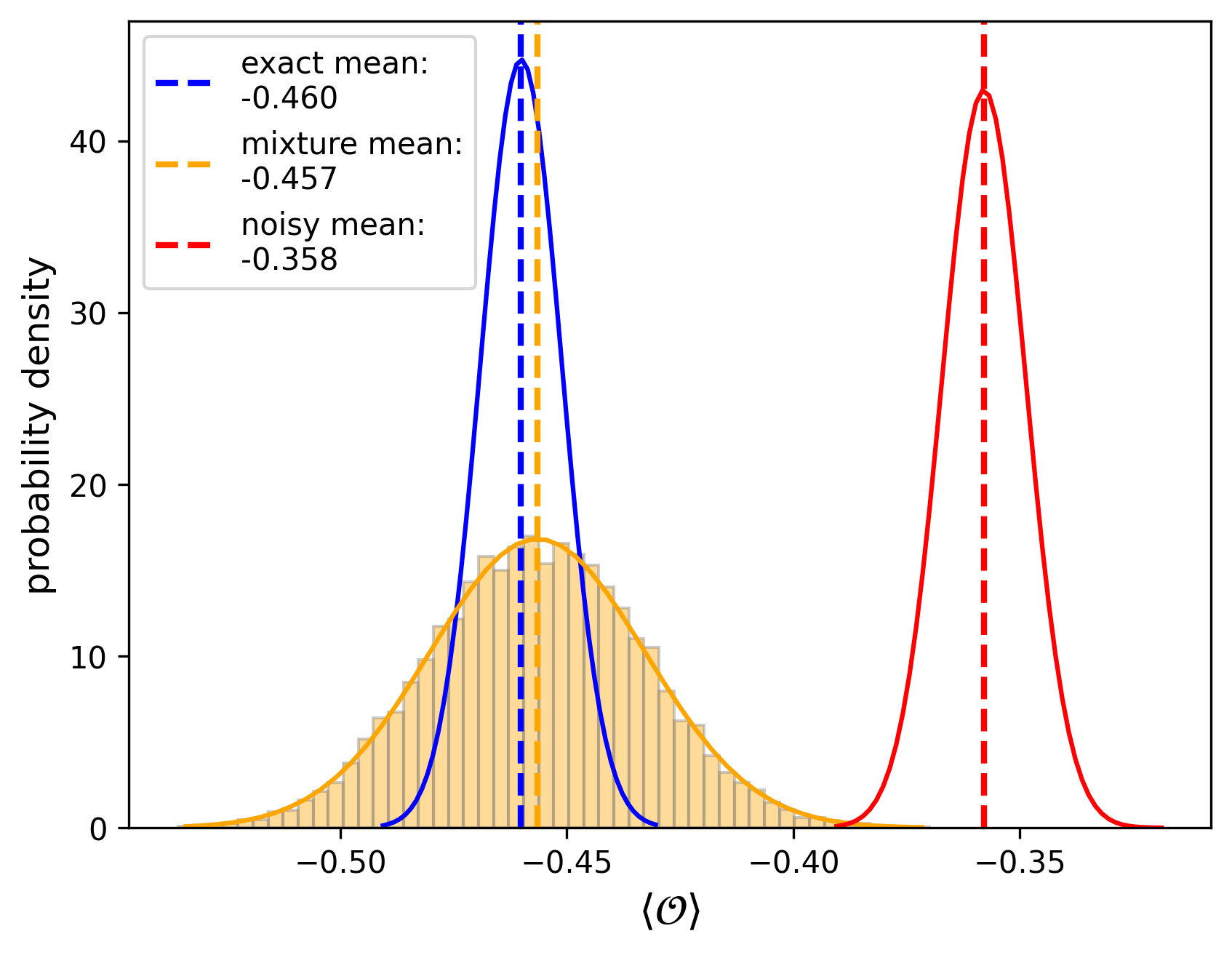}
    \caption{
    Estimating expectation value in a presence of a constant over-rotation $\epsilon=0.001$ along $Z$-direction at every gate of circuit defined in Eq.~\ref{eq:noisy_circuit}. The noisy expectation value (red dashed line) shows significant deviation from the exact value (blue dashed line). By applying the proposed method, the noiseless expectation value is effectively recovered, as indicated by the orange dashed line closely aligning with the blue one.
    }
\label{fig:errormitigationconstant}
\end{figure}

The results from the simulations are presented in Figure~\ref{fig:errormitigationconstant}.
The constant over-rotation error at each gate shifts the mean of the expected value distribution. 
However, using our method, we can recover the ideal expectation value. 
As anticipated, the variance of the distribution obtained through our method increases compared to the exact and noisy results. 
Note that the variance of distributions obtained with probabilistic mixtures appears higher than in Figure~\ref{fig:twirlingerrormitigation}, even though the effective noise strength $\epsilon = \sqrt{\epsilon_x^2 + \epsilon_y^2 + \epsilon_z^2} = 0.001$ was the same in Sec.~\ref{subsec:approx_error}. This is because, there, our method only mitigates the residual error in $Z$-direction $\epsilon_z \sim \epsilon / \sqrt{3}$.

The requirement of knowing the precise error parameters, as specified in Eq.~\ref{eq:noisy_circuit}, in order to apply our method may appear unrealistic, as such information could, in principle, be used to adjust control pulses and eliminate systematic over-rotations through calibration.
Nevertheless, in scenarios where coherent errors persist despite perfect calibration, we demonstrate that our method can still effectively mitigate these errors and yield accurate estimates of expectation values, provided that partial knowledge of the error parameters is available.
Additional examples demonstrating the applicability of our approach under such conditions are presented in Appendix\ref{ap:add_num_exp}.

\section{Conclusion}\label{sec:conclusion}
In this work, we presented a method to accurately estimate the expectation values of Hermitian observables from quantum circuits affected by unitary error channels. 
Our approach involves probabilistically inserting additional gates into the quantum circuit and combining their outputs according to a constructed unbiased estimator. 
This strategy effectively recovers the noiseless quantum channel, enabling precise estimation of target properties.

Numerical experiments validate our findings through two potential use cases.
The first focuses on suppressing unitary errors arising from the finite gate set decomposition of continuous rotations, demonstrating the method’s applicability in the fault-tolerant regime.
The second addresses unitary errors caused by imprecision in quantum hardware, a critical limitation in near-term quantum devices. 

While our method effectively suppresses unitary errors, the primary trade-off is the exponential growth in the number of shots required to achieve the desired accuracy. 
This scaling depends on both the error magnitude and the number of gates in the circuit, limiting the method’s applicability to systems with moderate error rates and circuit sizes. Nonetheless, our results indicate that for circuits of modest depth and error rates, the method reliably recovers accurate expectation values, underscoring its potential for error suppression in practical quantum computations.

Future research will explore the connections between our method and the classical simulatibility of quantum circuits. 
Additionally, we plan to extend our work to accurately characterize noise in physical devices.
Another key direction involves adapting our strategy for practical implementation on physical devices by tailoring the method to specific circuits, observables, and noise characteristics, ensuring its applicability in realistic environments.

\section*{Acknowledgments}
We thank Natalie Klco and Thomas Barthel for valuable discussions and insights on the manuscript.
This work was supported by the National Science Foundation (NSF) Quantum Leap Challenge Institute of Robust Quantum Simulation (QLCI grant OMA-2120757) and the NSF National Quantum Virtual Laboratory (NQVL) pilot program Quantum Advantage-Class Trapped Ion system (QACTI grant OSI-2410675).

\bibliography{main.bib}

\clearpage
\newpage
\onecolumngrid

\appendix
\section*{Appendix}\label{sec:Appendix}
\renewcommand{\thesubsection}{\Alph{subsection}}
\renewcommand{\theequation}{\thesubsection\arabic{equation}}

\subsection{Unitary decomposition}\label{ap:un_decomp}
\setcounter{equation}{0}
In this section, we derive the decomposition of a unitary rotation as a linear combination of two unitary gates.
Given a unitary gate, $R(\theta)=e^{-i\frac{\theta}{2}\mathbb{P}}$, where $\mathbb{P}$ is a Pauli-string and $\{\theta,\epsilon\}$ are real numbers, we can show the following:
\begin{align}
    R(\theta)&= e^{-i\frac{\theta}{2}\mathbb{P}} =  e^{-i(\frac{\theta + \epsilon - \epsilon}{2})\mathbb{P}} =  e^{-i(\frac{\theta + \epsilon}{2})\mathbb{P}} e^{i\frac{\epsilon}{2}\mathbb{P}}\nonumber\\
    &= [\cos(\frac{\theta+\epsilon}{2}) \mathbb{I} - i\sin(\frac{\theta+\epsilon}{2}) \mathbb{P}] [\cos(\frac{\epsilon}{2}) \mathbb{I} + i \sin(\frac{\epsilon}{2}) \mathbb{P}]\nonumber\\
    &= \cos(\frac{\epsilon}{2}) [\cos(\frac{\theta+\epsilon}{2}) \mathbb{I} - i\sin(\frac{\theta+\epsilon}{2}) \mathbb{P}] + i \sin(\frac{\epsilon}{2}) \mathbb{P} [\cos(\frac{\theta+\epsilon}{2}) \mathbb{I} - i\sin(\frac{\theta+\epsilon}{2}) \mathbb{P}]\nonumber\\ 
    &= \cos(\frac{\epsilon}{2}) [\cos(\frac{\theta+\epsilon}{2}) \mathbb{I} - i\sin(\frac{\theta+\epsilon}{2}) \mathbb{P}] + \sin(\frac{\epsilon}{2}) [ i\cos(\frac{\theta+\epsilon}{2}) \mathbb{P} + \sin(\frac{\theta+\epsilon}{2}) \mathbb{I}] \nonumber\\ 
    &= \cos(\frac{\epsilon}{2}) [\cos(\frac{\theta+\epsilon}{2}) \mathbb{I} - i\sin(\frac{\theta+\epsilon}{2}) \mathbb{P}] + \sin(\frac{\epsilon}{2}) [ i\sin(\frac{\theta+\epsilon+\pi}{2}) \mathbb{P} - \cos(\frac{\theta+\epsilon+\pi}{2}) \mathbb{I}] \nonumber\\ 
    &= \cos(\frac{\epsilon}{2}) [\cos(\frac{\theta+\epsilon}{2}) \mathbb{I} - i\sin(\frac{\theta+\epsilon}{2}) \mathbb{P}] - \sin(\frac{\epsilon}{2}) [\cos(\frac{\theta+\epsilon+\pi}{2}) \mathbb{I} - i \sin(\frac{\theta+\epsilon+\pi}{2}) \mathbb{P}]\nonumber\\
    &=\cos(\frac{\epsilon}{2}) e^{-i(\frac{\theta+\epsilon}{2})\mathbb{P}} - \sin(\frac{\epsilon}{2}) e^{-i(\frac{\theta+\epsilon+\pi}{2})\mathbb{P}} \nonumber\\
    &= \cos(\frac{\epsilon}{2}) R(\theta+\epsilon) - \sin(\frac{\epsilon}{2})R(\theta+\epsilon+\pi)
. \label{eq:ID0}
\end{align}

\subsection{Channel decomposition}\label{ap:chan_decomp}
\setcounter{equation}{0}
In this section, we derive the decomposition of a channel as probabilistic mixture of four different channels. 
We first derive some useful identities using Eq.~\ref{eq:ID0} before we find the desired decomposition:
\begin{align}
    R(\theta+\epsilon) &(.) R^{\dagger}(\theta+\epsilon+\pi) = R(-\frac{\pi}{2})R\left(\theta+\epsilon+\frac{\pi}{2}\right) (.) R^{\dagger}\left(\theta+\epsilon+\frac{\pi}{2}\right) R^{\dagger}(\frac{\pi}{2}) \nonumber\\
    &= [\cos(\frac{-\pi}{4})\mathbb{I} - i \sin(\frac{-\pi}{4}) \mathbb{P}] [R\left(\theta+\epsilon+\frac{\pi}{2}\right) (.) R^{\dagger}\left(\theta+\epsilon+\frac{\pi}{2}\right)] [\cos(\frac{\pi}{4})\mathbb{I} + i \sin(\frac{\pi}{4}) \mathbb{P}] \nonumber\\
    &= [\frac{1}{\sqrt{2}}\mathbb{I}+\frac{i}{\sqrt{2}}\mathbb{P}][R\left(\theta+\epsilon+\frac{\pi}{2}\right) (.) R^{\dagger}\left(\theta+\epsilon+\frac{\pi}{2}\right)][\frac{1}{\sqrt{2}}\mathbb{I}+\frac{i}{\sqrt{2}}\mathbb{P}] \nonumber\\
    &= \frac{1}{2}[R\left(\theta+\epsilon+\frac{\pi}{2}\right) (.) R^{\dagger}\left(\theta+\epsilon+\frac{\pi}{2}\right) -  \mathbb{P} R\left(\theta+\epsilon+\frac{\pi}{2}\right) (.) R^{\dagger}\left(\theta+\epsilon+\frac{\pi}{2}\right) \mathbb{P} \nonumber\\
    &+ i \mathbb{I}R\left(\theta+\epsilon+\frac{\pi}{2}\right) (.) R^{\dagger}\left(\theta+\epsilon+\frac{\pi}{2}\right) \mathbb{P} + i \mathbb{P} R\left(\theta+\epsilon+\frac{\pi}{2}\right) (.) R^{\dagger}\left(\theta+\epsilon+\frac{\pi}{2}\right) \mathbb{I} ].
\label{eq:ID1}
\end{align}
Similarly,
\begin{align}
    R(\theta+\epsilon+\pi) &(.) R^{\dagger}(\theta+\epsilon) = R(\frac{\pi}{2})R\left(\theta+\epsilon+\frac{\pi}{2}\right) (.) R^{\dagger}\left(\theta+\epsilon+\frac{\pi}{2}\right) R^{\dagger}(\frac{-\pi}{2}) \nonumber\\
    &= [\cos(\frac{\pi}{4})\mathbb{I} - i \sin(\frac{\pi}{4}) \mathbb{P}] [R\left(\theta+\epsilon+\frac{\pi}{2}\right) (.) R^{\dagger}\left(\theta+\epsilon+\frac{\pi}{2}\right)] [\cos(\frac{-\pi}{4})\mathbb{I} + i \sin(\frac{-\pi}{4}) \mathbb{P}] \nonumber\\
    &= [\frac{1}{\sqrt{2}}\mathbb{I}-\frac{i}{\sqrt{2}}\mathbb{P}][R\left(\theta+\epsilon+\frac{\pi}{2}\right) (.) R^{\dagger}\left(\theta+\epsilon+\frac{\pi}{2}\right)][\frac{1}{\sqrt{2}}\mathbb{I}-\frac{i}{\sqrt{2}}\mathbb{P}] \nonumber\\
    &= \frac{1}{2}[R\left(\theta+\epsilon+\frac{\pi}{2}\right) (.) R^{\dagger}\left(\theta+\epsilon+\frac{\pi}{2}\right) -  \mathbb{P} R\left(\theta+\epsilon+\frac{\pi}{2}\right) (.) R^{\dagger}\left(\theta+\epsilon+\frac{\pi}{2}\right) \mathbb{P} \nonumber\\
    &- i \mathbb{I}R\left(\theta+\epsilon+\frac{\pi}{2}\right) (.) R^{\dagger}\left(\theta+\epsilon+\frac{\pi}{2}\right) \mathbb{P} - i \mathbb{P} R\left(\theta+\epsilon+\frac{\pi}{2}\right) (.) R^{\dagger}\left(\theta+\epsilon+\frac{\pi}{2}\right) \mathbb{I} ].
\label{eq:ID2}
\end{align}

Using the identities in Eq.~\ref{eq:ID1} and Eq.~\ref{eq:ID2}, we derive the following:
\begin{align}
    R(\theta+\epsilon) &(.) R^{\dagger}(\theta+\epsilon+\pi) + R(\theta+\epsilon+\pi) (.) R^{\dagger}(\theta+\epsilon) \nonumber\\
    &= R\left(\theta+\epsilon+\frac{\pi}{2}\right) (.) R^{\dagger}\left(\theta+\epsilon+\frac{\pi}{2}\right) 
    -  \mathbb{P} R\left(\theta+\epsilon+\frac{\pi}{2}\right) (.) R^{\dagger}\left(\theta+\epsilon+\frac{\pi}{2}\right) \mathbb{P} \nonumber\\
    &= R\left(\theta+\epsilon+\frac{\pi}{2}\right) (.) R^{\dagger}\left(\theta+\epsilon+\frac{\pi}{2}\right)  + R(-\pi) R\left(\theta+\epsilon+\frac{\pi}{2}\right) (.) R^{\dagger}\left(\theta+\epsilon+\frac{\pi}{2}\right) R^{\dagger}(-\pi)\nonumber\\
    &= R\left(\theta+\epsilon+\frac{\pi}{2}\right) (.) R^{\dagger}\left(\theta+\epsilon+\frac{\pi}{2}\right)  + R\left(\theta+\epsilon-\frac{\pi}{2}\right) (.) R^{\dagger}\left(\theta+\epsilon-\frac{\pi}{2}\right).
 \label{eq:ID3}
\end{align}

Now, we can derive the decomposition of the channel $\mathcal{R}(\theta)$ by using the result of Eq.~\ref{eq:ID3}:
\begin{align}
    \mathcal{R}(\theta) &= R(\theta) (.) R^{\dagger}(\theta) \nonumber \\
     &= [\cos(\frac{\epsilon}{2}) R(\theta+\epsilon) - \sin(\frac{\epsilon}{2})R(\theta+\epsilon+\pi)] (.)  [\cos\left(\frac{\epsilon}{2}\right) R^{\dagger}(\theta+\epsilon) - \sin\left(\frac{\epsilon}{2}\right)R^{\dagger}(\theta+\epsilon+\pi)] \nonumber \\
     &= \cos^{2}\left(\frac{\epsilon}{2}\right) R(\theta+\epsilon) (.) R^{\dagger}(\theta+\epsilon) + \sin^{2}\left(\frac{\epsilon}{2}\right) R(\theta+\epsilon+\pi) (.) R^{\dagger}(\theta+\epsilon+\pi) \nonumber\\
     &- \cos\left(\frac{\epsilon}{2}\right)\sin\left(\frac{\epsilon}{2}\right)[R(\theta+\epsilon) (.) R^{\dagger}(\theta+\epsilon+\pi) + R(\theta+\epsilon+\pi) (.) R^{\dagger}(\theta+\epsilon) ] \nonumber\\
     &= \cos^{2}\left(\frac{\epsilon}{2}\right) \mathcal{R}(\theta+\epsilon) + \sin^{2}\left(\frac{\epsilon}{2}\right) \mathcal{R}(\theta+\epsilon+\pi) - \cos\left(\frac{\epsilon}{2}\right)\sin\left(\frac{\epsilon}{2}\right) [\mathcal{R}\left(\theta+\epsilon+\frac{\pi}{2}\right) - \mathcal{R}\left(\theta+\epsilon-\frac{\pi}{2}\right)].
\label{eq:ID4}
\end{align}

\subsection{Mixture of channels}\label{ap:mix_chan}
\setcounter{equation}{0}
In this section, we show the details of the probabilistic mixture used in the article.
First, we note that if we fix $\epsilon = -\theta$ in Eq.~\ref{eq:ID4}, we recover the following decomposition used in previous studies~\cite{Koczor2022,Koczor2023, kiumi2024te}:
\begin{align}
\mathcal{R}(\theta) &= \frac{1+\cos(\theta)}{2}I
+ \frac{\sin(\theta)}{2}[\mathcal{R}\left(\frac{\pi}{2}\right) - \mathcal{R}\left(-\frac{\pi}{2}\right)] + \frac{1-\cos(\theta)}{2}\mathcal{R}(\pi)
\label{eq:decomposition}
\end{align}
Then, we fix $\theta_1, \theta_2$ and $\theta_3$ in Eq.~\ref{eq:mixture} as:
\begin{align}
    \theta_1 &= \theta + \epsilon, \nonumber \\
    \theta_2 &= \theta + \epsilon + A, \nonumber \\
    \theta_3 &= \theta + \epsilon + B.
\end{align}
Then, using the combination of Eq.~\ref{eq:mixture} and Eq.~\ref{eq:decomposition}, we can get a linear system of equations for $\gamma_1, \gamma_2$, and $\gamma_3$:
\begin{equation}
\begin{bmatrix} 
1 + \cos(\theta + \epsilon) & 1 + \cos(\theta + \epsilon + A) & 1 + \cos(\theta + \epsilon + B) \\ 
\sin(\theta + \epsilon) & \sin(\theta + \epsilon + A) & \sin(\theta + \epsilon + B) \\ 
1 - \cos(\theta + \epsilon) & 1 - \cos(\theta + \epsilon + A) & 1 - \cos(\theta + \epsilon + B)
\end{bmatrix}
\begin{bmatrix} 
\gamma_1 \\ \gamma_2 \\ \gamma_3 
\end{bmatrix} = 
\begin{bmatrix} 
1 + \cos(\theta) \\ \sin(\theta) \\ 1 - \cos(\theta)
\end{bmatrix}.
\label{eq:matrix}
\end{equation}
The general solution for \eqref{eq:matrix} reads
\begin{align}
\gamma_1 &= \csc\left(\frac{A}{2}\right) \csc\left(\frac{B}{2}\right) \sin\left(\frac{A + \epsilon}{2}\right) \sin\left(\frac{B + \epsilon}{2}\right) \nonumber \\
\gamma_2 &= \csc\left(\frac{A}{2}\right) \csc\left(\frac{A - B}{2}\right) \sin\left(\frac{\epsilon}{2}\right) \sin\left(\frac{B + \epsilon}{2}\right) \nonumber \\
\gamma_3 &= -\csc\left(\frac{A - B}{2}\right) \csc\left(\frac{B}{2}\right) \sin\left(\frac{\epsilon}{2}\right) \sin\left(\frac{A + \epsilon}{2}\right).
\label{eq:generalsolution}
\end{align}

\subsection{Finding an optimal solution}\label{ap:opt_sol}
\setcounter{equation}{0}
We next use the general solution in Eq.~\ref{eq:generalsolution} and find a solution, i.e., values of $A$ and $B$, for which $\|\gamma\|_1=|\gamma_1| + |\gamma_2| + |\gamma_3|$ is minimized.
To do so, we solve the following problem:
\begin{equation}
    \text{min}(\|\gamma\|_1): \text{ subject to } A > B.
\end{equation}
We find that as $A \rightarrow 2\pi$ and $B \rightarrow \pi$, $\|\gamma\|_1$ is minimized.
Thus, in previous studies~\cite{Koczor2023, kiumi2024te}, the values of $A$ and $B$ are fixed to be $A=2\pi+\Delta$ and $B=\pi+\Delta$, where $\Delta$ is a small real number.

\begin{figure}[htbp!]
    \centering
    \includegraphics[width=0.9\textwidth]{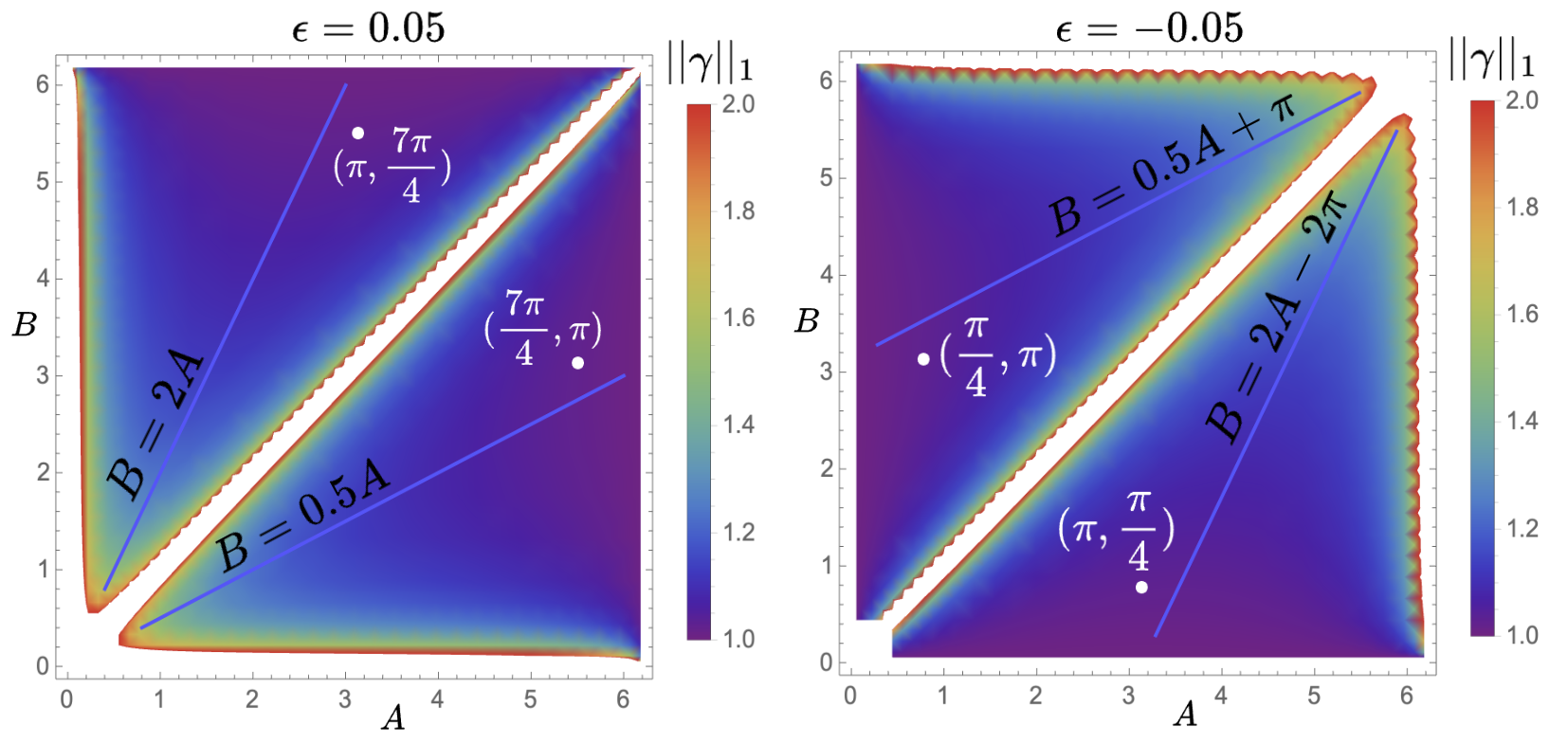}
    \caption{Projection of $\|\gamma\|_1$ onto $AB-$plane. 
    (left) Positive unitary error $\epsilon$. In the $A>B$ regime, for any chosen $A \in (0, 2\pi)$, the optimal $B$ is $0.5A$. The white dot in $A>B$ region corresponds to the decomposition in \ref{eq:ourmixture}, in which $A = -\frac{\pi}{4}$ and $B=\pi$. (right) Negative unitary error $\epsilon$. The plot is reflected across the diagonal $B=2\pi - A$ compared to the positive $\epsilon$ case. The optimal choice for $B$ becomes $2A-2\pi$ for any given $A$.}
    \label{fig:heatmap}
\end{figure}

We notice that for a given value of $A$, the optimal value of $B$ satisfies $B = 0.5A$ or $B = 2A$.
This can be seen from the plot in Figure~\ref{fig:heatmap}.
We also notice that there are many suboptimal solutions of $A$ and $B$ for which $\|\gamma\|_1$ is very close to the optimal solution.
Thus, we fix the values of $A$ and $B$ to be $A = 2\pi-\frac{\pi}{4} \equiv -\frac{\pi}{4} \text{ (mod } 2\pi), B = \pi$ and show that even for this suboptimal solution we can suppress the effect of unitary errors in sufficiently deep quantum circuits. These choices of $A$ and $B$ are motivated by the goal of minimizing the overhead associated with decomposing into the Clifford+$T$ gate sequence.

\subsection{Estimators}\label{ap:est}
\setcounter{equation}{0}
In this section, we show that the details of the estimators for the channels and expectation value mentioned in the paper.

\subsubsection{Over-rotation channel}
Given the estimator $\hat{\mathcal{R}}(\theta)$ in Eq.~\ref{eq:gateestimator} and the probability of choosing each $\mathcal{R}_i$ as $p_i = |\gamma_i(\epsilon)| / \|\gamma(\epsilon)\|_1$, we see that:
\begin{align}
    \mathbb{E}[\hat{\mathcal{R}}(\theta)]&= \sum_i p_i \hat{\mathcal{R}}(\theta) = \sum_i p_i \|\gamma(\epsilon)\|_1 \text{sign}[{\gamma_i(\epsilon)}]\mathcal{R}_i(\theta)\nonumber\\
    &= \sum_i \frac{|\gamma_i(\epsilon)|}{\|\gamma(\epsilon)\|_1} \|\gamma(\epsilon)\|_1 \text{sign}[{\gamma_i(\epsilon)}]\mathcal{R}_i(\theta)\nonumber\\
    &= \sum_i {\gamma_i(\epsilon)}\mathcal{R}_i(\theta) = \mathcal{R}(\theta).
\end{align}

\subsubsection{Twirled noisy channel}
Given the estimator $\hat{\mathcal{R}}_z(\theta)$ in Eq.~\ref{eq:CT_gateestimator} and the probability of choosing each $\bar{\mathcal{R}}_{z_{i}}^{\sigma_j}$ as $p_i = |\gamma_i(\epsilon_1)| / 4 \|\gamma(\epsilon_1)\|_1$, we can see that:
\begin{align}
    \mathbb{E}[\hat{\mathcal{R}}_z(\theta)]&= \sum_{i,j} p_i \hat{\mathcal{R}}_z(\theta) = \sum_{i,j} p_i \|\gamma(\epsilon_1)\|_1 \text{sign}[{\gamma_i}(\epsilon_1)] \bar{\mathcal{R}}_{z_{i}}^{\sigma_j}(\theta)\nonumber\\
    &= \sum_{i,j} \frac{|\gamma_i(\epsilon_1)|}{4 \|\gamma(\epsilon_1)\|_1} \|\gamma(\epsilon_1)\|_1 \text{sign}[{\gamma_i}(\epsilon_1)] \bar{\mathcal{R}}_{z_{i}}^{\sigma_j}(\theta) \nonumber\\
    &= \sum_i \gamma_i(\epsilon_1) \frac{1}{4} \sum_j \sigma_j \bar{R}_{z_{i}}(\theta) (.) \bar{R}_{z_{i}}^{\dagger} (\theta) \sigma_j^{\dagger}\nonumber\\
    &= \sum_i \gamma_i(\epsilon_1) \mathcal{T}_{R_{z_i}(\theta)} \nonumber \\
    &\approx \sum_i {\gamma_i(\epsilon_1)}\mathcal{R}_{z_i}(\theta) = \mathcal{R}_z(\theta).
\end{align}

\subsubsection{Circuits}\label{ap:est_circuits}
First, let's express $\mathcal{U}_c$ as a product of channels $\mathcal{R}^{(j)}$ to obtain an expression in terms of sampled instances $\mathcal{U}_L$:
\begin{align}
    \mathcal{U}_c = \prod_{j=1}^{\nu}\mathcal{R}^{(j)}(\theta_j)
    &= \prod_{j=1}^{\nu}[\gamma_1(\epsilon_j)\mathcal{R}_1^{(j)}+\gamma_2(\epsilon_j)\mathcal{R}_2^{(j)}+\gamma_3(\epsilon_j)\mathcal{R}_3^{(j)}] \nonumber \\
    &= \sum_L \prod_{j=1}^{\nu} \gamma_L^{(j)}(\epsilon_j) \mathcal{U}_L = \sum_L \Gamma_L \mathcal{U}_L,
\end{align}
where the sum runs over all $3^v$ possible circuits.

Now, given the estimator $\hat{\mathcal{U}}_c$ in Eq.~\ref{eq:circuit_estimator} and the probability of sampling a circuit configuration $\mathcal{U}_L$ as $p_L = |\Gamma_L|/\Gamma_c$, we can see that:
\begin{align}
    \mathbb{E}[\hat{\mathcal{U}}_c] &= \sum_L p_L \hat{\mathcal{U}}_c\nonumber\\
    &=  \sum_L p_L \Gamma_c \text{sign}[\Gamma_L] \mathcal{U}_L \nonumber \\
    &= \sum_L \frac{|\Gamma_L|}{\Gamma_c} \Gamma_c  \text{sign}[\Gamma_L] \mathcal{U}_L \nonumber \\
    &= \sum_L \Gamma_L \mathcal{U}_L \nonumber \\
    &= \sum_L \prod_{j=1}^{\nu} {\gamma_{l_j}^{(j)}{(\epsilon_j)}} \mathcal{R}^{(j)}_{l_j}(\theta_j) \nonumber \\
    &= \prod_{j=1}^{\nu}\mathcal{R}^{(j)}(\theta_j) = \mathcal{U}_c.
\end{align}

\subsubsection{Observables}
Given the estimator ${\hat{o}}$ as in Eq.~\ref{eq:exp_estimator}, the probability of observing an outcome $\ket{b}$ is $p_b = \text{ Tr}[E_b \mathcal{U}_L  \ketbra{0}{0}]$ and the probability of sampling a circuit $\mathcal{U}_L$, $p_L=|\Gamma_L|/\Gamma_c$. We can prove that the estimator is unbiased as follows:
\begin{align}
    \mathbb{E}[\hat{o}] &= \sum_{L,b} p_b \: p_L \: \hat{o} \nonumber\\
    &= \sum_{L,b} \text{ Tr}[E_b \mathcal{U}_L \ketbra{0}{0}] \frac{|\Gamma_L|}{\Gamma_c} \Gamma_c \text{sign}[\Gamma_L]  \text{Tr}[\mathcal{O}E_b] \nonumber \\
    &= \sum_{L,b} \Gamma_L \text{Tr}[\mathcal{O}E_b] \text{ Tr}[E_b \mathcal{U}_L \ketbra{0}{0}] \nonumber \\
    &= \sum_b \text{Tr}[\mathcal{O}E_b] \text{ Tr}[E_b(\sum_L \Gamma_L \mathcal{U}_L) \ketbra{0}{0}]\nonumber\\
    &= \sum_b \text{Tr}[\mathcal{O}E_b] \text{ Tr}[ E_b \mathcal{U}_c  \ketbra{0}{0}] \nonumber\\
    &= \text{ Tr}[(\sum_b \text{Tr}[\mathcal{O}E_b]E_b) \mathcal{U}_c \ketbra{0}{0}] = \text{ Tr}[\mathcal{O} \mathcal{U}_c \ketbra{0}{0}] =\braket{{O}}_{\mathcal{U}_c}.
\end{align}

\subsection{Sampling overhead}\label{ap:overhead}
\setcounter{equation}{0}
To calculate the bound on the number of shots, $S$, required to estimate the expectation value $\braket{\mathcal{O}}$ of an observable $\mathcal{O}$ within accuracy $\delta$, we first need to deduce the variance of its unbiased estimator $\hat{o}$ given in Formula \ref{eq:exp_estimator}. In this, we adapt the derivation from Ref.~\cite{Koczor2023}. First, note that $\text{Var}[\hat{o}] = \mathbb{E}[\hat{o}^2]-\mathbb{E}[\hat{o}]^2$ and that $\mathbb{E}[\hat{o}]^2 = \braket{\mathcal{O}}^2$ since the estimator is unbiased. So, $\text{Var}[\hat{o}]$ is bounded by $\mathbb{E}[\hat{o}^2]$:
\begin{align}
    \text{Var}[{\hat{o}}] &\le \mathbb{E}[\hat{o}^2] \nonumber \\
    &= \sum_{L, b}p_Lp_b \times (\Gamma_c \text{sign}[\Gamma_L] \text{Tr}[\mathcal{O}E_b])^2 \nonumber \\
    &= \Gamma_c^2 \sum_{L, b}p_Lp_b \times \text{Tr}[\mathcal{O}E_b]^2 \nonumber \\
    &\le \Gamma_c^2 \: \|\mathcal{O}\|_{\infty}^2,
\end{align}
where $p_L$ is the probability of sampling sequence $L = \left\{l_1...l_\nu\right\}$ and $p_b$ is the probability of measuring the outcome $\text{Tr}[\mathcal{O}E_b]$.

By definition, $\Gamma_c=\prod_{j=1}^{\nu} \|\gamma^{(j)}(\epsilon_j)\|_1$. Assuming over-rotations are equal at all gates ($\epsilon_j = \epsilon \: \forall \: j$), we write
\begin{align}
    \text{Var}[\hat{o}] \le \|\gamma(\epsilon)\|_1^{2\nu} \|\mathcal{O}\|_{\infty}^2.
\label{eq:est_variance}
\end{align}
For the choice of $A = -\text{sign}(\epsilon)\frac{\pi}{4}, B = \pi$ in Formula \ref{eq:generalsolution}, we obtain
\begin{align}
\|\gamma(\epsilon)\|_1 &= \sec \left(\frac{\pi }{8}\right) \cos \left(\epsilon-\frac{\pi }{8}\right) \nonumber \\
&= 1+\epsilon \tan \left(\frac{\pi }{8}\right)+O\left(\epsilon^2\right) \nonumber \\
&\approx 1 + 0.414\epsilon \nonumber \\
&\approx e^{0.414\epsilon}.
\end{align}
Substituting this into \ref{eq:est_variance}, we get 
\begin{align}
    \text{Var}[\hat{o}] &\le \|\gamma(\epsilon)\|_1^{2\nu}\|\mathcal{O}\|_{\infty}^2 \nonumber \\
    &\approx e^{0.83\epsilon\nu}\|\mathcal{O}\|_{\infty}^2.
\end{align}
Thus, the number of shots, $S$, required to estimate $\braket{\mathcal{O}}$ up to accuracy $\delta$ is
\begin{align}
    S &\le \text{Var}[\hat{o}]/\delta^2 \nonumber \\
    &\approx e^{0.83\epsilon\nu}\|\mathcal{O}\|_{\infty}^2 /\delta^2
\end{align}
as claimed in Statement \ref{statement}.

\subsection{Choosing the number of circuit instances}\label{ap:num_circuits}
In this appendix, we examine the dependence of expectation value estimation accuracy on the number of sampled noisy circuit instances, $N_c$. For this, we calculated $\braket{\mathbb{Z}^{\otimes N}}$ after $30$-Trotter-step time evolution under Hamiltonian in Eq.~\ref{eq:hamiltonian} on $N=12$ qubits using $S=10^5$ shots.
Assuming a constant over-rotation of $\epsilon = 0.01$ at each gate (as in Sec.~\ref{subsec:coherent_or}), we varied $s$ between $1$ and $5\times 10^4$, thereby sampling different numbers of noisy circuits to construct the observable estimator (Eq.~\ref{eq:exp_estimator}).

\begin{figure}[htbp!]
    \centering
    \includegraphics[width=0.5\textwidth]{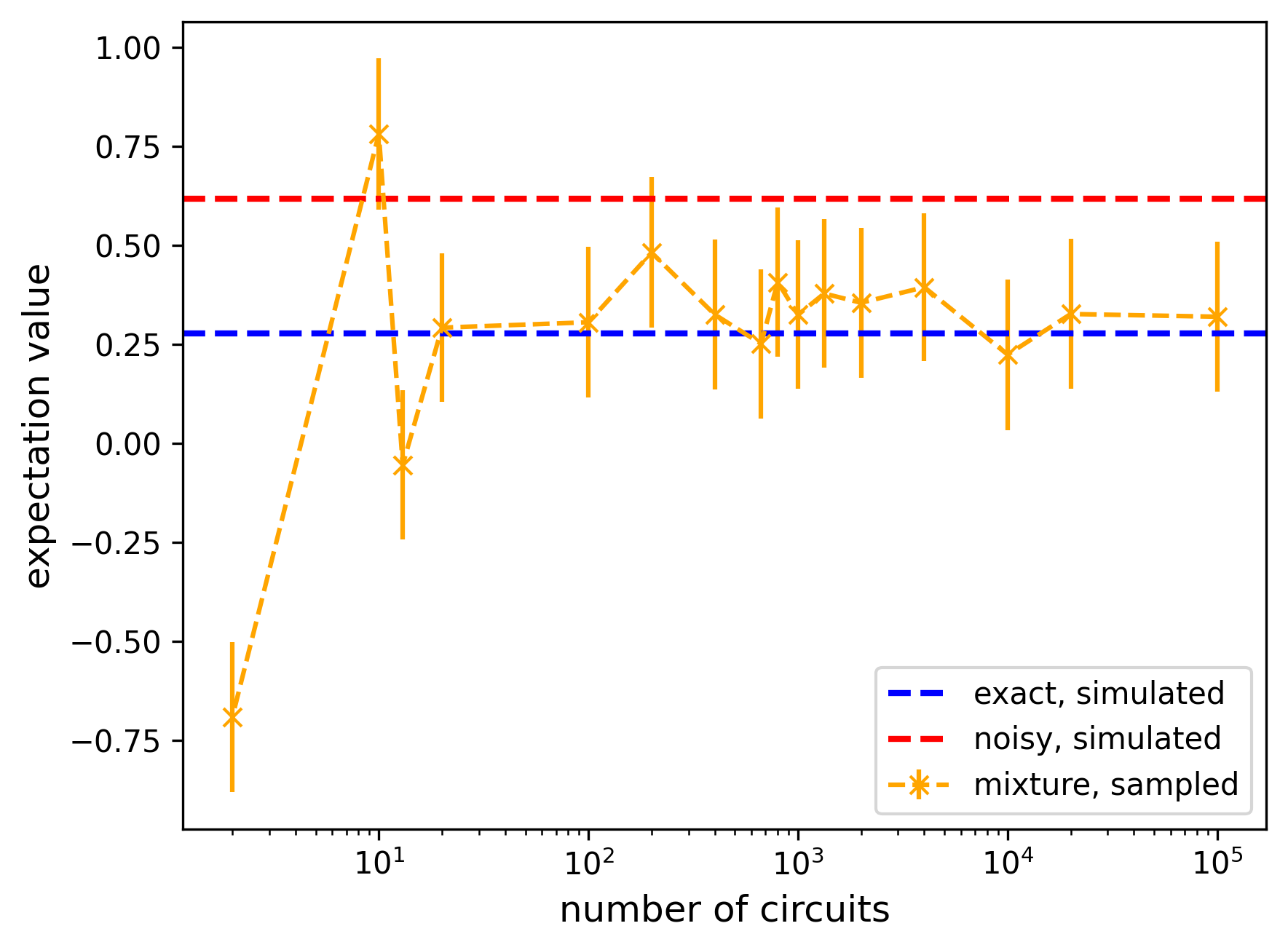}
    \caption{Dependence of the estimated expectation value on the number of sampled circuit instances used in the probabilistic mixture of the circuit estimator (Eq.~\ref{eq:circuit_estimator}). As the number of sampled circuits increases, the estimator converges, and the improvement over the unmitigated noisy estimate becomes apparent.}
    \label{fig:numberofcircuits}
\end{figure}

The simulation results are presented in Figure~\ref{fig:numberofcircuits}.
As expected, using too few circuit variations in our method leads to poor expectation value estimates, usually performing worse than the noisy calculation. 
This effect is particularly evident at the data point corresponding to just two circuit instances. 
Additionally, note that the uncertainty in the obtained quantity remains independent of $N_c$, as it is solely determined by the circuit size and over-rotation magnitude (see Appendix\ref{ap:overhead} for details).

Based on these results, we fixed the number of shots per sampled circuit, $s$, to $100$ in all the numerical simulations described in Sec.~\ref{sec:results}.
Thus, the number of sampled circuit instances is $N_c = S \times 10^{-2}$, where $S$ is the total number of shots we possess to calculate the expectation value $\braket{\mathcal{O}}$.

\subsection{Visualizing the results}\label{ap:visualizing}
\setcounter{equation}{0}
For a given simulation, we obtain an array of $S$ numbers, each corresponding to a measurement. In the noiseless and noisy circuits, these measurements are binary, taking values of -1 or +1. In our protocol, these values are weighted by circuit prefactors as defined in Eq.~\ref{eq:circuit_estimator}. To plot the histogram, we uniformly sample $10^4$ results from the array and calculate their average. This process is repeated $10^4$ times, resulting in $10^4$ expectation value estimates. These estimates are then grouped into 50 bins and presented in Figures~\ref{fig:twirlingerrormitigation}, \ref{fig:errormitigationconstant}, and \ref{fig:errormitigationuniform}. 

\subsection{Circuits}\label{ap:circuitry}
\setcounter{equation}{0}
\subsubsection{Hamiltonian simulation}
\begin{figure}[htbp!]
    \centering
    \includegraphics[width=0.3\textwidth]{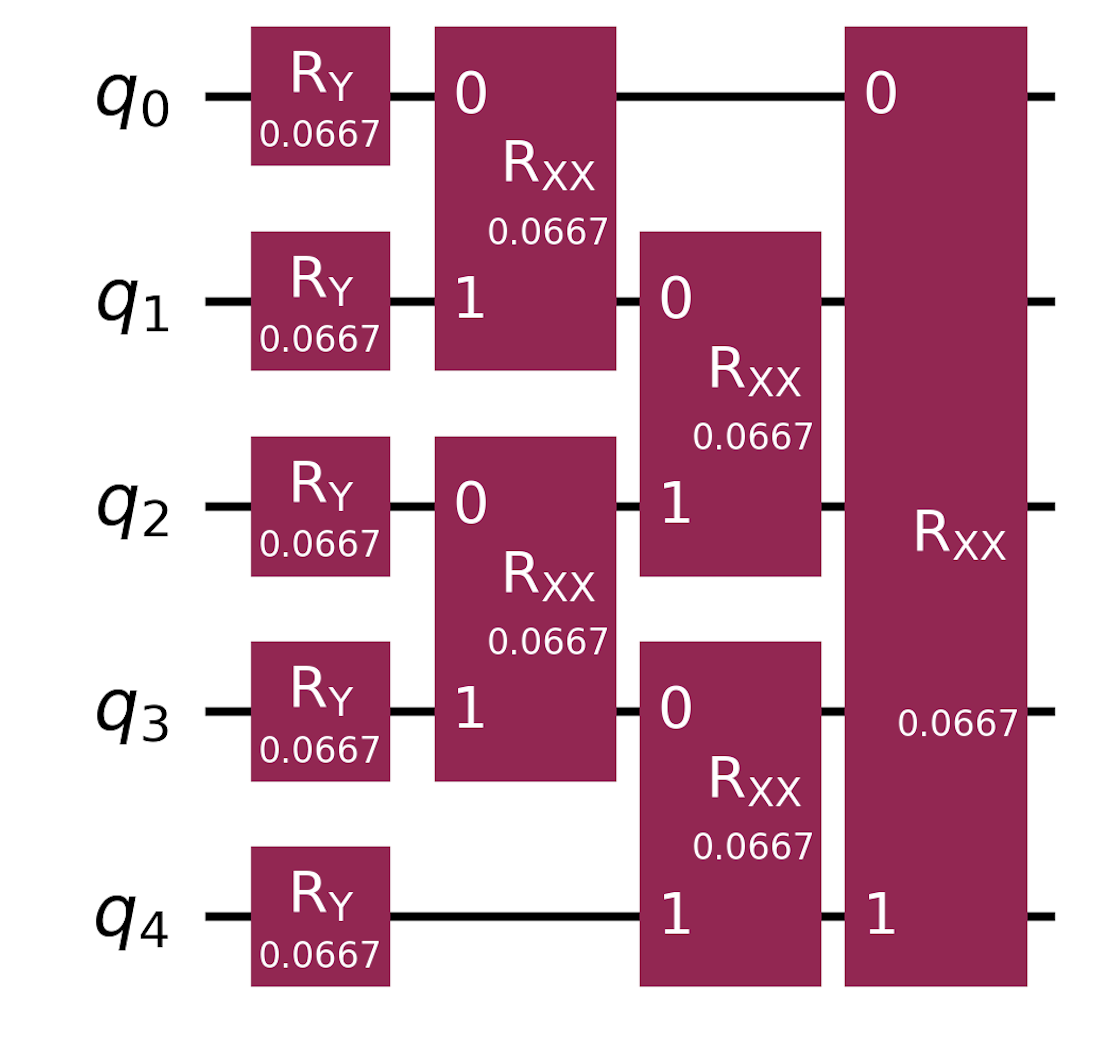}
    \caption{A single layer of a time evolution circuit for 5 qubits. The layer of single-qubit $R_y$ rotations is followed by $XX-$ interactions. Periodic boundary conditions impose the connection between the first qubit ($q_0$) and the $N^{\text{th}}$ qubit ($q_4$).}
    \label{fig:singlelayer}
\end{figure}

A single layer of a 30-layered time evolution circuit under Hamiltonian in Eq.~\ref{eq:hamiltonian} with $N=5$ is shown in Figure~\ref{fig:singlelayer}. There are $2N = 10$ parametrized gates in total. Note that all rotations are through the angle $2T/\mathcal{L} = 2/30$ since we chose $h=J=1$.

\subsubsection{Compilation in FT era}
\begin{figure}[htbp!]
    \centering
    \includegraphics[width=1\textwidth]{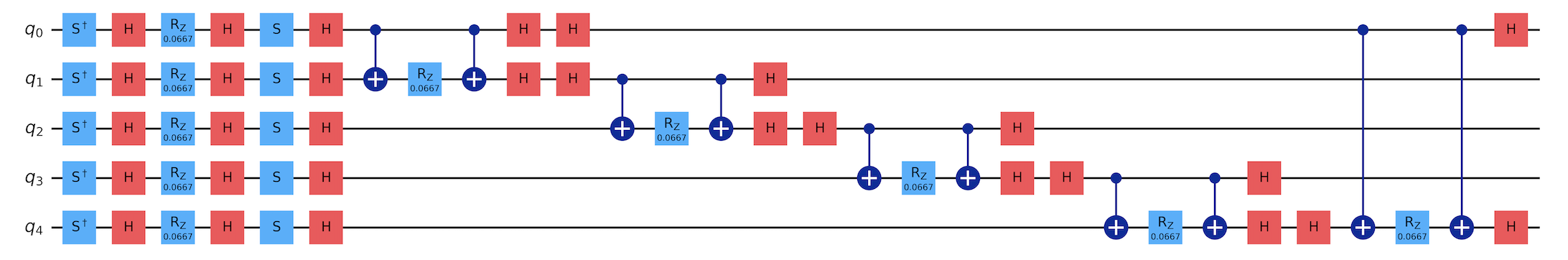}
    \caption{A decomposed single layer of a 5-qubit time evolution. $R_{y}(\theta)$ is mapped to $\mathbb{SH}R_{z}(\theta)\mathbb{HS}^\dagger$. $R_{\mathbb{X}_i\mathbb{X}_j}(\theta)$ is mapped to $\mathbb{H}_i\mathbb{H}_j\text{CNOT}_{ij} R_{z_j}(\theta)\text{CNOT}_{ij}\mathbb{H}_i\mathbb{H}_j$. This circuit is equivalent to the one in Figure~\ref{fig:singlelayer}.}
    \label{fig:decomposedlayer}
\end{figure}

Before inserting Pauli twirls and rotations through $\pm\frac{\pi}{4}$ and $\pi$, we decomposed the $R_{y}$ and $R_{\mathbb{XX}}$ gates using $\{\text{CNOT}, \mathbb{H}, \mathbb{S}, \mathbb{S}^\dagger\}$. The decomposed 5-qubit layer is illustrated in Figure~\ref{fig:decomposedlayer}.

\begin{figure}[htbp!]
    \centering
    \includegraphics[width=1\textwidth]{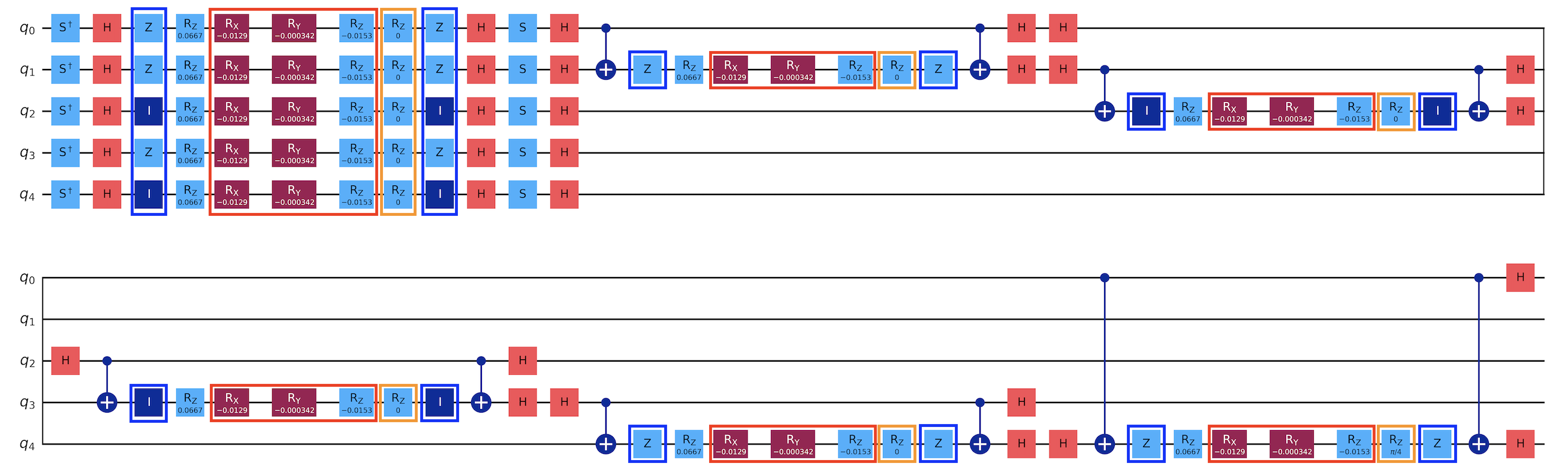}
    \caption{A sampled decomposed single layer of a 5-qubit time evolution with inserted gates. 
    Red boxes frame the added noise. Note that $\epsilon_z$ is negative in this case, so our strategy requires the insertion $\mathbb{I}$, $T$, and $\mathbb{Z}$. 
    Orange boxes frame the gates used for our method (rotations through $0, \frac{\pi}{4}$, and $\pi$).
    Blue boxes frame Pauli twirls ($\mathbb{I}$ and $\mathbb{Z}$).}
    \label{fig:layerinstance}
\end{figure}

An example of a twirled sampled circuit is in the Figure~\ref{fig:layerinstance}.

\subsection{Additional numerical experiments}\label{ap:add_num_exp}

\subsubsection{Constant over-rotation}\label{subap:const_or}

\begin{figure*}[htbp!]
    \centering
    \begin{tabular}{c c}

    \qquad (a) & \qquad (b) \\
    \includegraphics[width=0.48\textwidth]{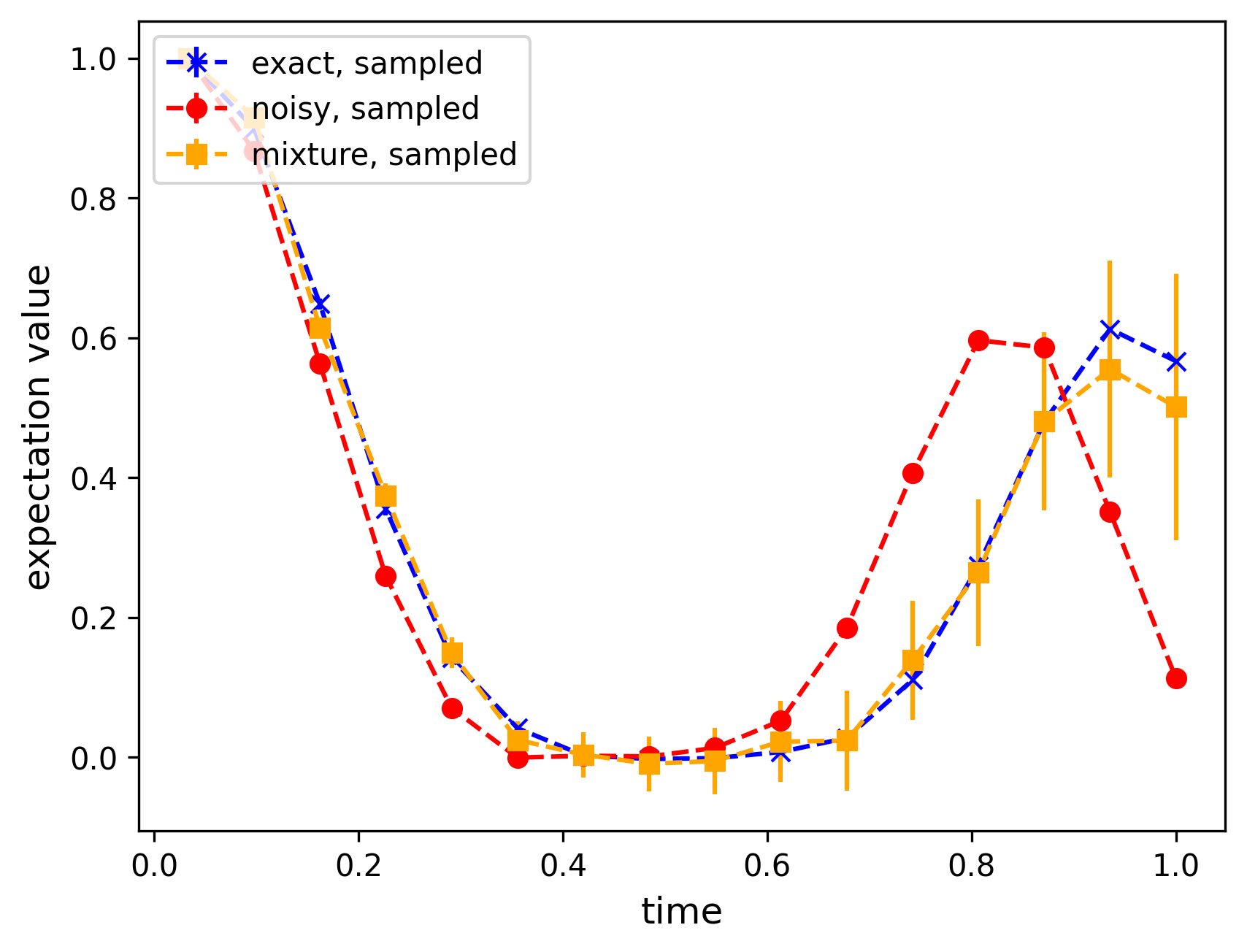} & 
    \includegraphics[width=0.5\textwidth]{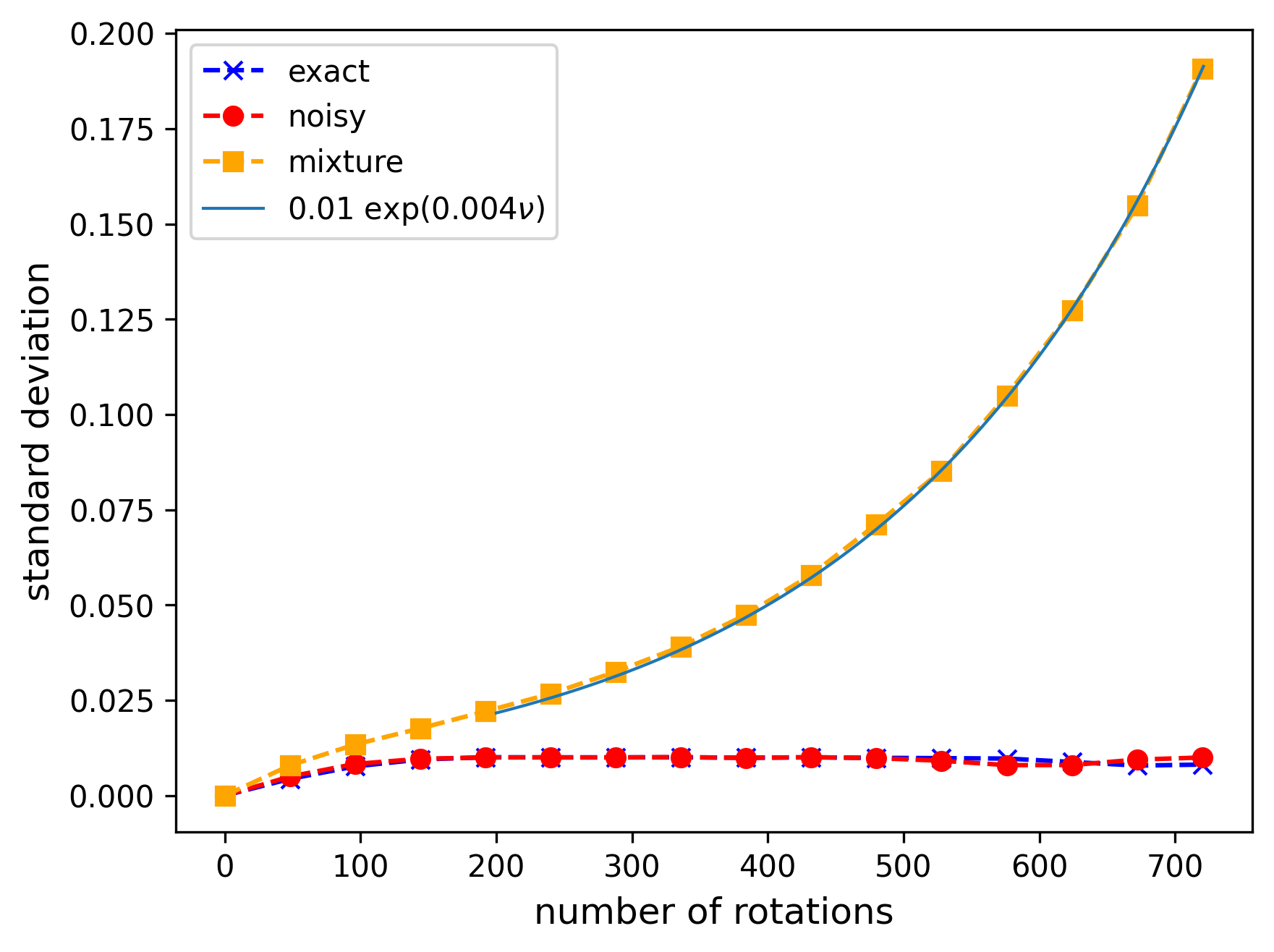} \\
    
    \end{tabular}
    \caption{
    (a) Time evolution of $\braket{\mathbb{Z}^{\otimes N}}$ under high coherent error ($\epsilon = 0.01$). The proposed method effectively mitigates noise, bringing the estimate obtained using probabilistic mixtures of noisy circuits (orange curve) closer to the exact value (blue curve) compared to the uncorrected noisy result (red curve).
    (b) Scaling of the standard deviation of the expectation value as the number of  parametrized  gates in the circuit grows. Due to our method, the scaling is exponential (as predicted by Eq.~\ref{eq:shotsscaling}), while the shape of the expectation value distribution is preserved for both exact and noisy circuits.
    } 
\label{fig:stepbystep}
\end{figure*}

In addition to the experiment described in Sec.~\ref{subsec:coherent_or}, we perform simulations to track the evolution of the expectation value $\braket{\mathcal{O}}$ after each Trotter step.
This allows us to analyze the behavior of our method as the number of gates in the circuit increases. 
For this, we use a Hamiltonian (Eq.~\ref{eq:hamiltonian}) on $N=12$ qubits and simulate the evolution up to $\mathcal{L}=30$ Trotter steps. 
The value of the over-rotation error is fixed at $\epsilon=0.01$ for this simulation.
The results from this numerical experiment are plotted in Figure~\ref{fig:stepbystep}~(a). Our method allows for significant improvement over the noisy case (orange curve, unlike the red curve, is closely aligned with the blue one). The noise reduction comes at the cost of an exponentially increasing standard deviation, as shown by rapidly widening error bars of orange data points. This happens because the number of shots is kept at the constant value $S=10^5$ as the quantum circuit deepens.
The standard deviation of the expectation value at time $T=1$ estimated using our method is appreciably higher than that shown in Figure~\ref{fig:errormitigationconstant} (0.19 vs. 0.02), despite the circuit having less  parametrized  gates. This is because $\epsilon$ is an order of magnitude lower in the latter simulation (0.01 vs. 0.001).

We study the growth in the standard deviation of $\braket{\mathcal{O}}$ as a function of the number of circuit parameters in Figure~\ref{fig:stepbystep}~(b).
As predicted by Eq.~\ref{eq:shotsscaling}, the variance of the expectation value distribution increases exponentially with the growing circuit size when the number of shots, $S$, remains constant. To manage this, $S$ must scale according to Eq.~\ref{eq:shotsscaling}. 
Note that the spread in expectation value does not depend on the circuit size for the exact and noisy cases. 

Finally, we highlight the difference in the evolution of the expectation value between Figure~\ref{fig:stepbystep}~(a) and Figure~\ref{fig:twirledmixturestepbystep}.
In Figure~\ref{fig:stepbystep}~(a), systematic over-rotation by $\epsilon$ effectively accelerates the time evolution, causing the red curve to lead the blue curve. 
In contrast, in Figure~\ref{fig:twirledmixturestepbystep}, less structured coherent errors result in noisy expectation values that fluctuate around the exact value, with red data points deviating either above or below the blue ones.

\subsubsection{Probabilistic over-rotation}
In most cases, the exact value of the over-rotation error $\epsilon$ is typically unknown.
Instead, one usually has an approximate value accompanied by a standard deviation.
In this subsection, we consider the case where the over-rotation error in each gate is sampled uniformly from an interval $[-\epsilon_0, 3\epsilon_0]$, and we use the mean value $\epsilon_0$ for calculating coefficients in Eq.~\ref{eq:ourgammas} at each gate.

In this numerical experiment, we fix $N=15$ in the Hamiltonian (Eq.~\ref{eq:hamiltonian}) and simulate the time evolution using $\mathcal{L}=70$ trotter steps utilizing $S=3\times 10^6$ shots for calculating the expectation value.
We begin with estimating $\braket{\mathcal{O}}$ using an exact (noiseless) circuit.
We then estimate $\braket{\mathcal{O}}$ using the noisy circuit in Eq.~\ref{eq:noisy_circuit}. Unlike in Sec.~\ref{subsec:coherent_or}, the over-rotations are now different at each gate and are sampled from the uniform distribution. Finally, we apply the estimator in Eq.~\ref{eq:exp_estimator} to evaluate $\braket{\mathcal{O}}$, assuming $\epsilon=\epsilon_0$.
The total number of circuits $S/s=3\times10^4$ were constructed according to the methods in Sec.~\ref{sec:mix_noisy_circuits}, each subject to different over-rotation parameters.

\begin{figure}[htbp!]
    \centering
    \includegraphics[width=0.50\textwidth]{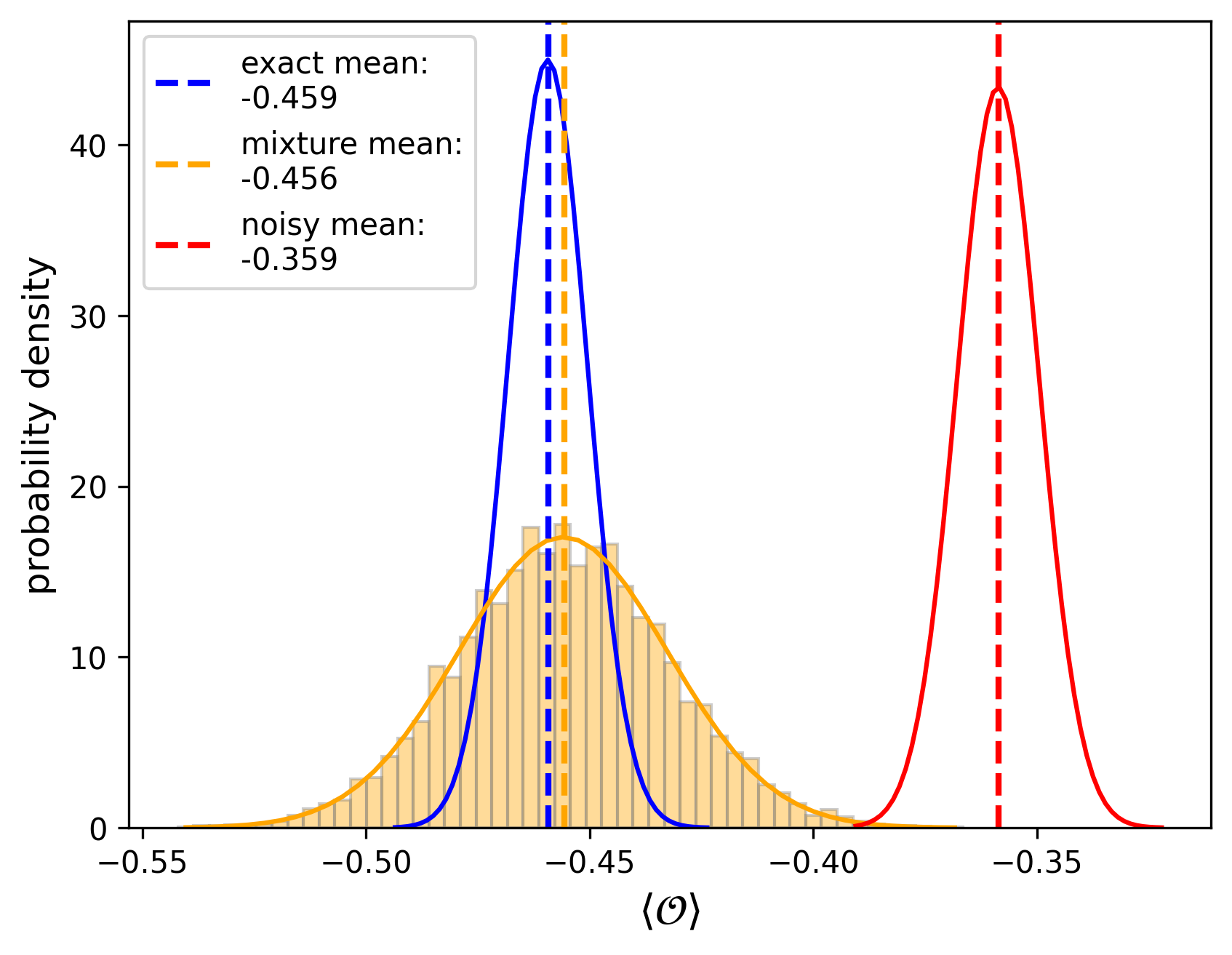}
    \caption{
    Estimating the expectation value $\braket{\mathcal{O}}$ in the presence of coherent over-rotations along $Z$-direction uniformly sampled from $[-\epsilon_0, 3\epsilon_0]$ at each gate of the circuit defined in Eq.\ref{eq:noisy_circuit}. The noisy expectation value (red dashed line) shows a significant deviation from the exact value (blue dashed line). By applying the proposed method, the noiseless expectation value is effectively recovered, as indicated by the orange dashed line closely aligning with the blue one. The assumption $\epsilon = \epsilon_0 = 0.001$ is used to construct the estimators in Eq.\ref{eq:exp_estimator}.
    }
    \label{fig:errormitigationuniform}
\end{figure}

The results from these simulations are presented in Figure~\ref{fig:errormitigationuniform}.
The coherent over-rotation error $\epsilon$ sampled uniformly from $[-\epsilon_0, 3\epsilon_0]$ at each gate shifts the mean of the expected value distribution. 
Notably, our method remains effective even with randomness in $\epsilon$, provided the mean error is known precisely.
As anticipated, the variance of the distribution obtained using our method increases compared to the exact and noisy results.

\end{document}